\documentclass[useAMS,usenatbib]{mn2e}

\usepackage{graphicx,amsmath,eufrak}
\usepackage{natbib}
\usepackage{fixltx2e}
\usepackage{hyperref}

\newcommand{\msun}{M$_{\sun}$}

\newcommand\tiri{{\sc TiRiFiC}}
\newcommand\HI{H{\sc i}}

\newcommand\Halpha{H$\alpha$ }

\title[HALOGAS observations of NGC 5023 and UGC 2082]{HALOGAS observations of NGC 5023 and UGC 2082: Modeling of non-cylindrically symmetric gas distributions in edge-on galaxies}
\author[P. Kamphuis et al. ]{P. Kamphuis$^{1,2,3}$\thanks{E-mail: peter.kamphuis@csiro.au}, R. J. Rand$^{4}$,  G. I. G. J\'ozsa$^{5,6}$, L. K. Zschaechner$^{4}$, G. H. Heald$^{5}$,  \newauthor M. T. Patterson$^{7}$\thanks{Visiting Astronomer, Kitt Peak National Observatory, National Optical Astronomy Observatory, which is operated by the Association of Universities for Research in Astronomy (AURA) under cooperative agreement with the National Science Foundation.},  G. Gentile$^{8,9}$, R. A. M. Walterbos$^7$, P. Serra$^5$ and W. J. G. de Blok$^{5,10}$ \\
$^{1}$CSIRO Astronomy \& Space Science, P.O. Box 76, Epping, NSW 1710, Australia\\
$^{2}$Astronomisches Institut Ruhr-Universit\"at Bochum, Universit\"atstrasse 150, D-44801 Bochum, Germany\\
$^{3}$Humboldt Fellow\\
$^{4}$Department of Physics and Astronomy, University of New Mexico, 1919 Lomas Blvd NE, Albuquerque, New Mexico, USA\\
$^{5}$ASTRON, the Netherlands Institute for Radio Astronomy, Postbus 2, 7990 AA, Dwingeloo, The Netherlands \\
$^{6}$Argelander-Institut f\"ur Astronomie, Auf dem H\"ugel 71, D-53121 Bonn\\
$^{7}$Department of Astronomy, New Mexico State University, PO Box 30001, MSC 4500, Las Cruces, NM 88003, USA\\
$^{8}$Sterrenkundig Observatorium, Ghent University, Krijgslaan 281 S9, B-9000 Ghent, Belgium\\
$^{9}$Department of Physics and Astrophysics, Vrije Universiteit Brussel, Pleinlaan 2, 1050 Brussels, Belgium\\
$^{10}$Astrophysics, Cosmology and Gravity Centre, Department of Astronomy, University of Cape Town, Private Bag X3, Rondebosch 7701, South Africa}
\begin{document}
\date{}
\pagerange{\pageref{firstpage}--\pageref{lastpage}} \pubyear{2013}
\maketitle
\label{firstpage}
\begin{abstract}
In recent years it has become clear that the vertical structure of disk galaxies is a key ingredient for understanding galaxy evolution. In particular, the presence and structure of extra-planar gas has been a focus of research. The Hydrogen Accretion in LOcal GAlaxieS (HALOGAS) survey aims to provide a census on the rate of cold neutral gas accretion in nearby galaxies as well as a statistically significant set of galaxies that can be investigated for their extra-planar gas properties.\\   
\indent In order to better understand the the vertical structure of the neutral hydrogen in the two edge-on HALOGAS galaxies NGC 5023 and UGC 2082 we construct detailed tilted ring models. The addition of distortions resembling arcs or spiral arms significantly improves the fit of the models to these galaxies. In the case of UGC 2082 no vertical gradient in rotational velocity is required in either symmetric models nor non-symmetric models to match the observations.  The best fitting model features two arcs of large vertical extent that may be due to accretion.  In the case of NGC 5023 a vertical gradient is required in symmetric models  (dV/dz =$-14.9\pm3.8$ km s$^{-1}$ kpc$^{-1}$) and its magnitude is significantly lowered when non-symmetric models are considered (dV/dz =$-9.4\pm3.8$ km s$^{-1}$ kpc$^{-1}$). Additionally it is shown that the underlying disk of NGC 5023 can be made symmetric, in all parameters except the warp, in non-symmetric models. In comparison to the ``classical'' modeling these models fit the data significantly better with a limited addition of free parameters.  
\end{abstract}
\section{Introduction}
In the last decade it has become clear that extra-planar gas is common in spiral galaxies \citep{2003A&A...406..505R,Sancisi2008}.  Understanding the physics that govern this extra-planar gas as well as the disk-halo connection in these galaxies has become an important part of understanding how galaxies evolve. Currently the average cold neutral gas accretion rate is estimated at 0.2 M$_{\odot}$ yr$^{-1} $ for spiral galaxies  \citep{Sancisi2008}. It is difficult to relate this to the amount of gas that is converted into stars because the current star formation rate (SFR) in spiral galaxies varies throughout their disks as well as from galaxy to galaxy. However, considering the median SFR of 3  M$_{\odot}$ yr$^{-1} $ in the local universe \citep{2011MNRAS.415.1815B} most  galaxies would consume all their gas in $\sim$ 1 Gyr were it not replenished, possibly through the accretion of hot or cold  gas \citep{2005MNRAS.363....2K}, or the  consumption time extended through recycling \citep{Kennicut1994}. The halo of a galaxy will form its connection with the Intergalactic medium (IGM) and is the location where gas that is brought up from  the disk through galactic fountains  \citep{1976ApJ...205..762S,1980ApJ...236..577B} or chimneys \citep{1989ApJ...345..372N} resides. Therefore, it is clear that it plays a significant role in galaxy evolution.\\
\indent In this paper we will adopt the following definitions when discussing the distribution of hydrogen gas in galaxies. As standard practice we define $z$ as the distance from the galaxy mid plane and refer to the \HI\ density profile along the $z$ axis as the vertical profile. We define the main \HI\ disk as the gas component with the narrowest vertical profile which can be described by a function of $\frac{z}{z_0}$, where $z_0$ is the characteristic scale height. All gas in excess of this main disk will be referred to as extra-planar gas (or halo gas). Several different extra-planar gas components have been identified in the literature and we refer to them in the following manner. Whenever the smooth vertical gas distribution requires a second characteristic scale height to adequately describe the data, and this gas is rotating coherently with the main disk, this second component will be called a thick disk\footnote{Note that previously this component has also been referred to as a \HI\ gaseous halo. We prefer not to use this term to avoid confusion with the common understanding of halos as roughly spherical mass distributions.}. We would like to stress here that this definition does not imply or require hydrostatic equilibrium of this thick disk. Distinct elongated structures outside the main disk will be referred to as filaments or streams. These could originate from the disk as well as from outside the galaxy. Finally, there are extra-planar clouds, e.g. the high and intermediate velocity clouds in our own Milky Way \citep{1963CRAS..257.1661M,1997ARA&A..35..217W} or the counter rotating clouds in NGC 891 \citep{Oosterloo2007}.\\
\indent The number of spiral galaxies which are observed to the required sensitivity for detecting extra-planar gas, in \HI\ as well as other wavelengths, has steadily grown over the past years  \citep{1997ApJ...491..140S, 2000A&A...356L..49S,2001A&A...377..759L,2002AJ....123.3124F,2005A&A...439..947B,2005A&A...436..101W,2008A&A...490..555B, Zschaechner2011}. First and foremost there is  the very thick \HI\ disk around NGC 891 \citep{1997ApJ...491..140S, Oosterloo2007}.  This disk contains about 30\% of the total \HI\ mass of this galaxy and shows up as a distinct component in the vertical profile of the gas \citep{Oosterloo2007}. Possibly  one of the most notable features in the neutral gas  of NGC 891 is the  decline of its rotational velocities as a function of vertical distance to the plane \citep{2005ASPC..331..239F}, the so-called ``lag''. This lag is determined to be linear with a magnitude of  $\sim$-16 km s$^{-1}$ kpc$^{-1}$ in the vertical direction and confirmed in the ionized gas \citep{2006ApJ...647.1018H,Kamphuis2007a}.\\
\indent This lag has become an important observational diagnostic for understanding the physics of the disk-halo interface. Simulations have shown that the inferred lags cannot be explained by purely ballistic models of galactic fountains alone \citep{2002ApJ...578...98C,2007ApJ...663..933H,2008MNRAS.386..935F} -- these underpredict the observed lags. There are several theoretical possibilities that could remedy this discrepancy. Firstly, if the clouds that are expelled from the disk interact with an existing hot halo that has sufficiently slow rotation, it may absorb the angular momentum of the clouds and thus explain the observed steep vertical rotational velocity gradients \citep{2011MNRAS.415.1534M}. Even more so, the cold gaseous clouds traveling through the hot halo could be crucial in cooling the hot halo gas, thus replenishing the gas supply in the disk. Alternatively, the accreted cold gas might flow down onto the disk along the angular momentum axis in a cylindrical fashion naturally causing the observed vertical rotational velocity gradients \citep{2006MNRAS.370.1612K}. And lastly, it has been shown that, under the right conditions, the gradients could even exist in static distributions  \citep{2002ASPC..276..201B, 2006A&A...446...61B} while in fountain models, the effects of radial pressure gradients and magnetic tension could affect lags.\\
\indent Observations indicate that the extra-planar gas is a combination of accreted gas and gas originating from the disk. The low metallicity of several High Velocity Clouds (HVCs) surrounding the Milky Way \citep{2004ASSL..312..195V} and the presence of filaments and other irregular extra-planar gaseous structures in nearby galaxies \citep{Sancisi2008},  imply that the extra-planar gas must partially come from gas that has been accreted from the IGM and/or companion galaxies.  On the other hand, the fact that  in some nearby galaxies the extra-planar \HI\ is concentrated towards the disk \citep{Oosterloo2007}  or that the majority of high velocity clouds are located near the star forming inner disk \citep{2005ASPC..331..247B}, implies that a large part of this gas originates in the disk. However, the relative contributions as well as the underlying physics remain ill-understood.\\
\indent  A lag has now been observed in several  galaxies besides NGC 891  \citep{2007ApJ...663..933H,Zschaechner2012}. However, the number of galaxies where this lag is actually quantified remains limited \citep[see][for a recent overview]{Zschaechner2012}. As the vertical gradients are mostly quantified in edge-on galaxies their determinations are subject to severe line of sight effects \citep{1979A&A....74...73S}. One of the major uncertainties in the research up to now is that non-cylindrically symmetric distributions in or above the plane, besides any possible warping or flaring of the disk, have not been considered. This is a serious concern as it is well known that star formation predominantly takes place  in the spiral arms. Therefore, if the extra-planar gas is pushed out of the plane through galactic chimneys created by  the massive stars it is natural to expect that the extra-planar gas is also preferentially located above the arms. There are indications that such an effect takes place in the ionized gas of  NGC 891 \citep{Kamphuis2007b}. \\
\indent  Because the dynamics of the extra-planar gas  are to a large degree governed by gravity an interesting subset of galaxies to consider are smaller galaxies. There is observational evidence that in smaller galaxies ($v_{{\rm rot}} \leq 120$ km s$^{-1}$) the potential becomes too weak to collapse the dust component into a well defined thin disk \citep{2004ApJ...608..189D}, i.e thinner than the stellar disk. This indicates that the complex interplay between gravity and other physics governing the evolution of galaxies undergoes a significant change at galaxies of this characteristic size. Results from \HI\  studies in such small galaxies have varied significantly.  \cite{2003ApJ...593..721M} found tentative evidence for a significant lagging thick disk in UGC 7321 whereas NGC 4244 displays a lag without a clear extra-planar component \citep{Zschaechner2011}. And in UGC 1281   no extra-planar gas nor lag are required \citep{Kamphuis2011}, albeit in this last  case a lag cannot be completely excluded. These results clearly show that more detailed investigations into the \HI\ distribution in lower mass galaxies are required before they can be treated as a significant subsample of galaxies.\\
 \indent With the continued development of the Tilted Ring Fitting Code (\tiri) \citep{Joshtirific2007} asymmetric distributions\footnote{ A full description of the current version of {\sc TiRiFiC}, as well as the latest release, can be found at \url{http://www.astron.nl/~jozsa/tirific/}} and their effects on the apparent vertical gradient can now be explored. We will do this here through comparison with deep \HI\  observations of the  small edge-on  galaxies NGC 5023 and UGC 2082.  These observations were done as part of the Westerbork Hydrogen Accretion in LOcal GAlaxieS (HALOGAS) survey \citep{2011A&A...526A.118H} on the Westerbork Synthesis Radio Telescope (WSRT).  HALOGAS has observed 22 spiral galaxies to sufficient depth (typical column densities of $\sim$ a few times 10$^{18}$ cm$^{-2}$ at typical linewidths in normal spiral galaxies) to detect extra-planar gas, of which a significant subsample (12 galaxies) is of lower dynamical mass. It provides some of the deepest line-emission observations ever performed on spiral galaxies and  aims to provide a consensus on the current cold gas accretion rate as well as detect any correlation between the SFR and extra-planar gas in nearby galaxies.  \\
\indent Even though, at their assumed distances, UGC 2082 is more than twice the physical size of NGC 5023 (See Table \ref{tab0} for their basic parameters), both have a fairly low, but non negligible, SFR. However due to their low potential this could still lead to an extra-planar component when gas is ejected through SF processes.  
Although small galaxies are not expected to accrete baryonic matter at lower redshifts \citep{2010AdAst2010E..87H}, both NGC 5023 and UGC 2082 are in the transition region between accreting and non-accreting galaxies, and therefore a modest amount of lagging extra-planar gas might be present if the vertical gradient is predominantly formed by the accretion of matter. \\ 
\indent This article is structured as follows. In $\S$ \ref{Obs}  we will describe the data reduction and observations. $\S$ \ref{ModelsN5023} will contain the models for NGC 5023 and   $\S$ \ref{ModelsU2082}  for UGC 2082.  We will discuss our results in $\S$ \ref{disc} and provide a summary in $\S$ \ref{sum}.
\begin{table}
    \centering
   
    \begin{tabular}{@{} llllll @{}} 
       \hline
       \hline
	Parameter& NGC 5023 & UGC2082 & Ref.\\
       \hline
       	Morphological Type &Scd  & Scd &2\\
	Center ($\alpha$ J2000)&13$^{{\rm h}}$  12 $^{{\rm m}}$11.83$^{{\rm s}}$& $2^{\rm h} 36^{\rm m} 16.6^{\rm s}$& 1\\
	 \hspace*{0.99cm}($\delta$ J2000)&44$^{\circ}$2\arcmin\ 16.9\arcsec&$25^{\circ} 25\arcmin\  20\arcsec$ & 1\\	
	v$_{\rm sys}$ (km s$^{-1})$&405& 705& 1\\              		
 	Distance (Mpc) &6.6 &14.4 & 3\\
	M$_{\rm B}$ &17.29 &18.55 & 3\\
	D$_{25}$ (arcmin)&6.8&5.8&3\\
	D$_{25}$ (kpc)&13.1&24.3& 3\\
	D$_{\rm{H \textsc{i}}}$ (arcmin)$^{\rm{a}}$&9.7&10.5 &1\\
	D$_{\rm{H \textsc{i}}}$ (kpc)$^{\rm{a}}$&18.6&43.8&1\\		
	Total \HI\ Mass (M$_{\odot}$) & $6.1 \times 10^8$ & $2.54 \times 10^9$&1\\  
	v$_{\rm max}$ (km s$^{-1}$)&89&98&1\\
	SFR (M$_{\odot}$ yr$^{-1}$) &0.039&0.041& 4\\
 \hline
    \end{tabular}
     \caption{(1) This work, (2) \citet{1992yCat.7137....0D}, (3) \citet{2011A&A...526A.118H} and references therein, (4)  \citet{2012A&A...544C...1H}.
     a) The diameter of the galaxy at a column density of $10^{20}$ cm$^{-2}$}
    \label{tab0}
 \end{table}
\section{Observations and Data Reduction}\label{Obs}
\subsection{21 cm Line Observations}
The 21 cm line, or \HI, observations were obtained with the WSRT as a part of the HALOGAS program \citep{2011A&A...526A.118H}.  After the standard HALOGAS reduction and calibration \citep{2011A&A...526A.118H} additional analysis was performed with the {\sc gipsy} package \citep{1992ASPC...25..131V}. The final cube of NGC 5023 (UGC 2082) was reduced to 69 (72) velocity channels and was offline Hanning smoothed which resulted in velocity resolution of 4.12 km s$^{-1} $. For the analysis of the neutral hydrogen, two cubes with different spatial resolution were used. One cube at a high resolution of  FWHM = 19\arcsec\ $\times$ 13\arcsec\ (FWHM = 29\arcsec\ $\times$ 13\arcsec) and one at a lower resolution and almost circular beam of  FWHM = 36\arcsec\ $\times$ 33\arcsec\ (FWHM = 41\arcsec\ $\times$ 33\arcsec). This results in physical resolutions of 0.6 $\times$ 0.4 kpc (2.0 $\times$ 0.9 kpc) and  1.2 $\times$ 1.1 kpc (2.9 $\times$ 2.3 kpc) for the high and low resolution cubes, at given distance. In this paper only the high resolution cube is displayed, unless noted otherwise. However, results presented here are always checked for inconsistencies against the lower resolution cubes.\\
\subsection{Individual Features}
Even though this paper deals primarily with modeling the global structures of NGC 5023 and UGC 2082 we would like to point out several individual  features that are visible in the data.  The reason for this is twofold. Firstly, cataloguing these features, some of which are potential tracers of cold gas accretion, helps to obtain a clear picture of the cold gas accretion in the local universe.The second reason is that these features are localized deviations on top of the smooth underlying disk. Therefore it is important to ignore these features when trying to fit global models to the data. In order to do this consistently the individual features need to be identified. 
\subsubsection{NGC 5023}\label{N5023ind}
\begin{figure*}
   \centering
   \includegraphics[width=16 cm]{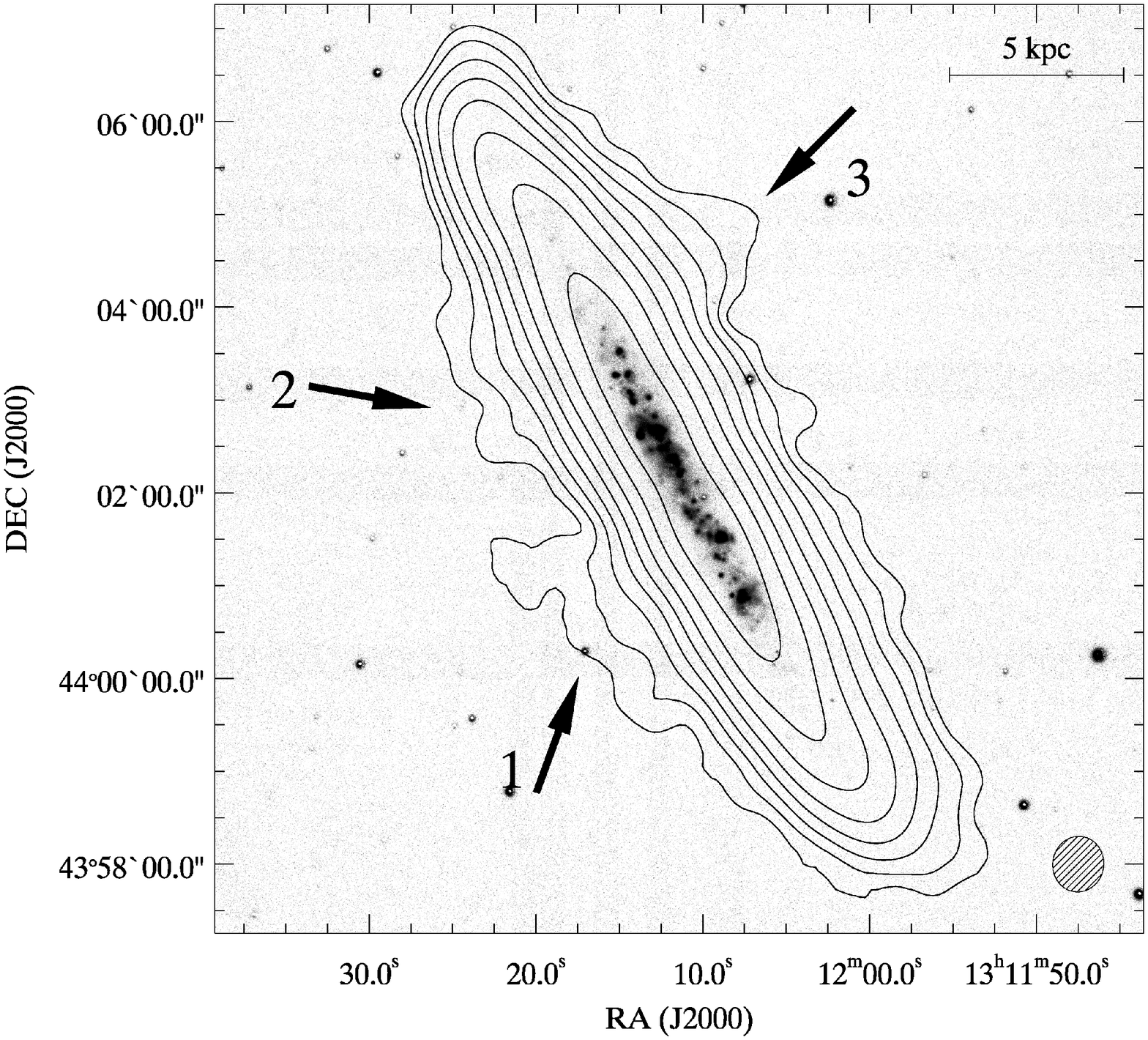} 
   \caption{NGC 5023: Contours of the low resolution \HI\ overlaid on an  \Halpha\  image (see text) of NGC 5023.  The contour levels are $2^n
 \times 10^{19}$\ cm$^{-2}$, $n=1-8$.  The arrows point out the features discussed in the text and the beam is shown in the lower right corner (36\arcsec\ $\times$ 33\arcsec)  $3\sigma = 8.3\times10^{18}$ cm$^{-2}$ for a width of 16.48 km s$^{-1}$ (4 channels).  }
   \label{mom0HA}
\end{figure*}

NGC 5023 shows three large distinct \HI\  features. These features are clearly visible in the low resolution zeroth moment map shown in Figure \ref{mom0HA}, as pointed out by the arrows. This map is made with the assumption of a negligible optical depth in the \HI\ observations. The underlying greyscale shows a 30 min \Halpha\ exposure from the Mosaic 1.1 Imager on the  4m Mayall Telescope  at Kitt Peak National Observatory (Patterson et al. in prep). Two of the features are located on the South East side (negative vertical offset) of the disk on opposite sides of the center. One is seen over the radial range +25\arcsec\ to +100\arcsec\  (South East) at vertical offsets of $\sim$ -75\arcsec\ (Fig \ref{mom0HA}, Arrow 1) and the other further from the center at radial offsets of -150\arcsec\ to -50\arcsec\ (Fig \ref{mom0HA}, Arrow 2) but at a similar vertical offset. The third feature can be seen on the North West side of the galaxy opposite to the second feature at positive vertical offsets (Fig \ref{mom0HA}, Arrow 3). All three features are coherent structures in velocity space spanning a range of roughly 60 km s$^{-1}$.  These parameters are summarized in Table \ref{tab2}. Unfortunately we were unable to obtain individual masses for these features as they are too integrated with the disk. We will return to the possible origin of these features in the discussion, but will initially ignore them in the modeling. \\
\indent It is interesting to note that the features do not seem to extend beyond the edge of the \Halpha disk (although feature 3 is right at the edge), which may imply a connection with the underlying star formation.  \cite{1996ApJ...462..712R} detected an extra-planar ionized hydrogen region in the quadrant of feature 2, however this ionized gas is offset (in projection) by 1.5\arcmin (2.9 kpc) from this feature. The two brightest \Halpha regions in the southern part of the disk, underlying feature 1, show diffuse extensions in the image presented here.  The extension around the most southern region is notable because it resembles an hourglass shape. These correlations with the \Halpha\ emission are interesting to note but mean little by itself. A multi-wavelength analysis would be necessary to understand the correlation with the SF in the disk. As such an analysis is beyond the scope of this paper we refer the reader to a future paper where this analysis will be done to the full extent.\\
\indent 
\subsubsection{UGC 2082}\label{U2082ind}
\begin{figure*}
\centering
\includegraphics[width= 14 cm]{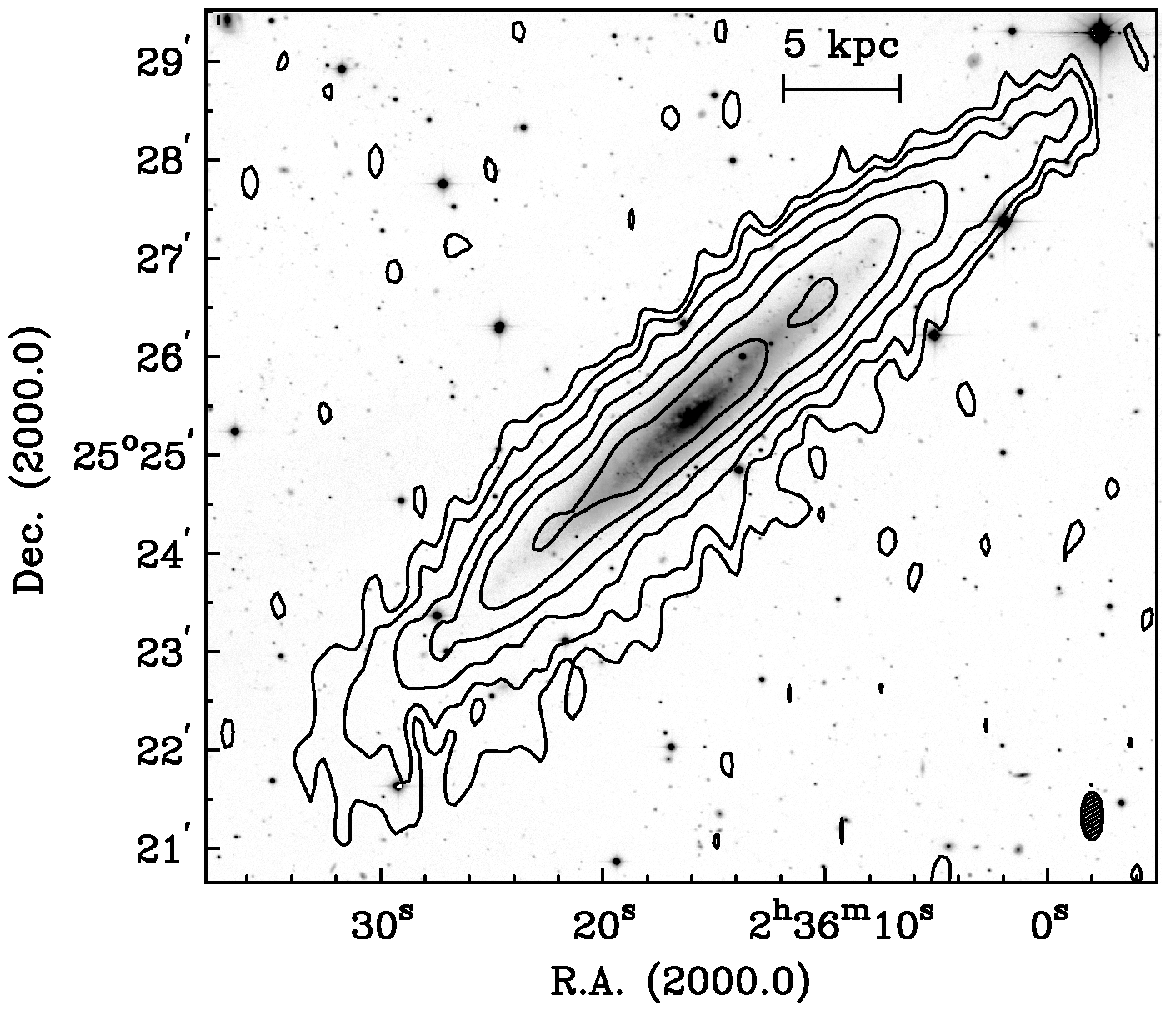}
\caption{UGC 2082: Contours of \HI\ overlaid on an {\it
r'}-band image (see text) of UGC 2082.  The contour levels are $2^n
\times 5.1 \times 10^{19}$\ cm$^{-2}$, $n=1-6$.  The beam is shown in
the lower right corner  (29\arcsec\ $\times$ 13\arcsec).  $3\sigma = 4.3\times10^{19}$ cm$^{-2}$ for a width of 16.48 km s$^{-1}$ (4 channels).}
\label{fig0}
\end{figure*}

\begin{figure}
   \centering
   \includegraphics[width=8 cm]{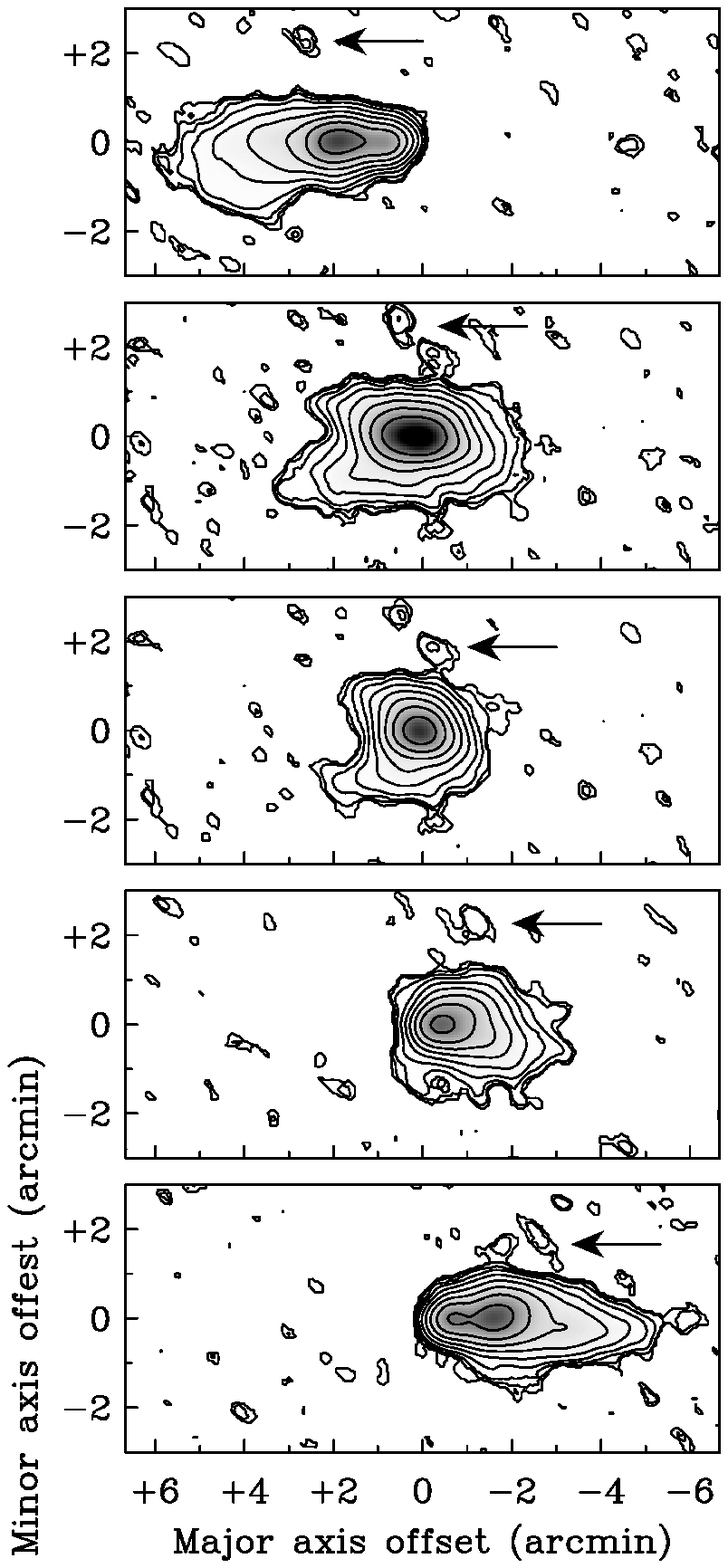} 
   \caption{UGC 2082:  zeroth moment maps of the five features listed in Table 2 as seen in the full-resolution cube (top to bottom: Features $1-5$).  Limited velocity ranges have been used to isolate the features: they are $688-712$ km s$^{-1}$, $667-724$ km s$^{-1}$, $721-741$ km s$^{-1}$,$622-647$ km s$^{-1}$, and $750-778$ km s$^{-1}$ for Features 1-5, respectively. }
   \label{fig0a}
\end{figure}

Figure \ref{fig0} shows \HI\ contours from our full resolution map overlaid on an {\it r'}-band image taken with the
Isaac Newton Telescope as part of the HALOSTARS project associated with HALOGAS.  The exposure time of the image is 2398 s (see Gentile et al. 2013\nocite{Gentile2013} for additional details). UGC 2082 was not detected in the deep H$\alpha$ survey of \cite{2003A&A...406..505R}.
\indent Five features, all above the N side of the disk, were found in the  low resolution cube
with 4$\sigma$ peaks in at least two consecutive channels.  Their
parameters are listed in Table \ref{tab2}.  None is 
counterrotating with respect to the disk emission below it.  It is difficult to estimate sizes for these features
as they are barely resolved.  If their distances from the minor axis
represent true vertical distances above the plane, then these range
from about 7 to 11 kpc. The fifth feature suggests a broken shell
of diameter $\sim$4 kpc at the adopted distance. The features as they appear in 
the full-resolution cube are shown in Figure \ref{fig0a}.
\begin{table*}
    \centering
    \begin{tabular}{@{} llllll @{}} 
       \hline
       \hline
       Feature   & Major axis &Minor axis &Mass &$V_0$ &$\Delta V_{FWZI}$ \\
  & distance (arcsec) & distance (arcsec) & ($10^5$ \msun) &(km s$^{-1}$) &(km s$^{-1}$)\\
       \hline
       \multicolumn{6}{c}{NGC 5023} \\ 
       
       \hline
       1 & 72\arcsec& -80\arcsec\ &  & 380 & 56\\
	2 & -112\arcsec\ & -92\arcsec\ &  & 444 & 40\\
	3 & -120\arcsec\ & 76\arcsec\ &  & 444 & 52\\
	 \multicolumn{6}{c}{(1 kpc = 31.25\arcsec)} \\ 
            \hline
       \multicolumn{6}{c}{UGC 2082} \\ 
      
       \hline
1 & 156\arcsec& 112\arcsec\ & 5.8 & 638 & 25\\
2 & 28\arcsec\ & 160\arcsec\ & 13 & 712 & 58\\
3 & -16\arcsec\ & 112\arcsec\ & 9.6 & 700 & 24\\
4 & -72\arcsec\ & 116\arcsec\ & 9.8 & 729 & 20\\
5 & -140\arcsec\ & 92\arcsec\ & 10 & 758 & 28\\
  \multicolumn{6}{c}{(1 kpc = 14.3\arcsec)} \\ 
\hline
    \end{tabular}
    \caption{Features found in the low resolution cube for NGC 5023 and UGC 2082. $V_0$ is the central velocity and $\Delta V_{FWZI}$ the Full Width at Zero Intensity of the specific feature.}
    \label{tab2}
 \end{table*}
\section{NGC 5023 Models}\label{ModelsN5023}
\begin{figure}
   \centering
   \includegraphics[width=9 cm]{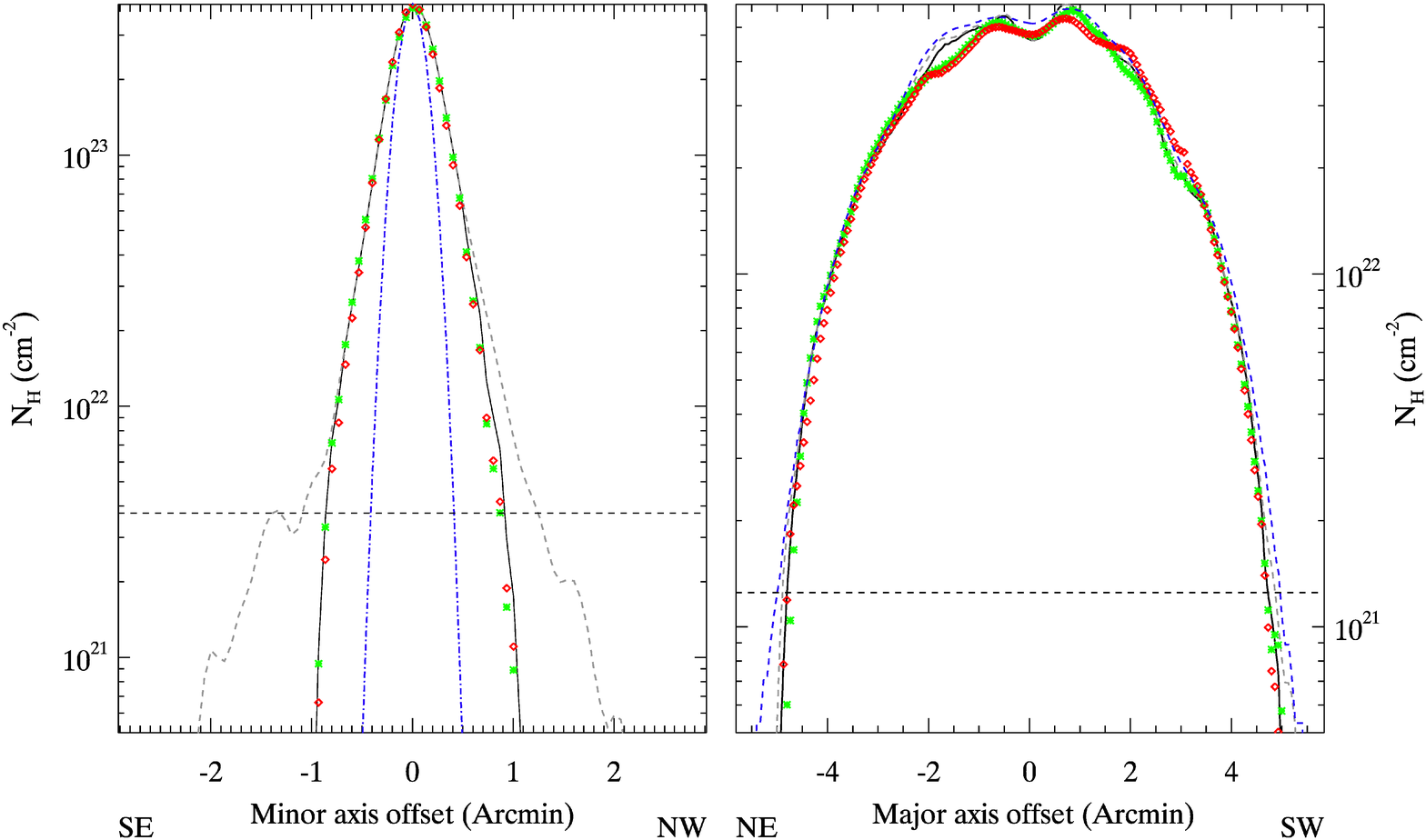} 
   \caption{NGC 5023: Left: minor axis profiles of the data (grey dashed line), of the data without the areas where the residual is higher than 50\%(solid black line),  Base model (Green asterisks) and the best fit Spiral Arms model (Red diamonds) summed over the inner 300\arcsec\ in the radial direction. Right: same as the left but now major axis profiles integrated over the inner 100\arcsec\ in the vertical direction. The blue dashed line shows the profile derived by the {\sc radial} task in {\sc gipsy} in the right panel and a Gaussian with the FWHM of the beam in the left panel. The horizontal black dashed lines indicate the lowest contour of figure \ref{mom0Model}  }
   \label{lineprofileHI}
\end{figure}

\begin{figure*}
   \centering
   \includegraphics[width=18 cm]{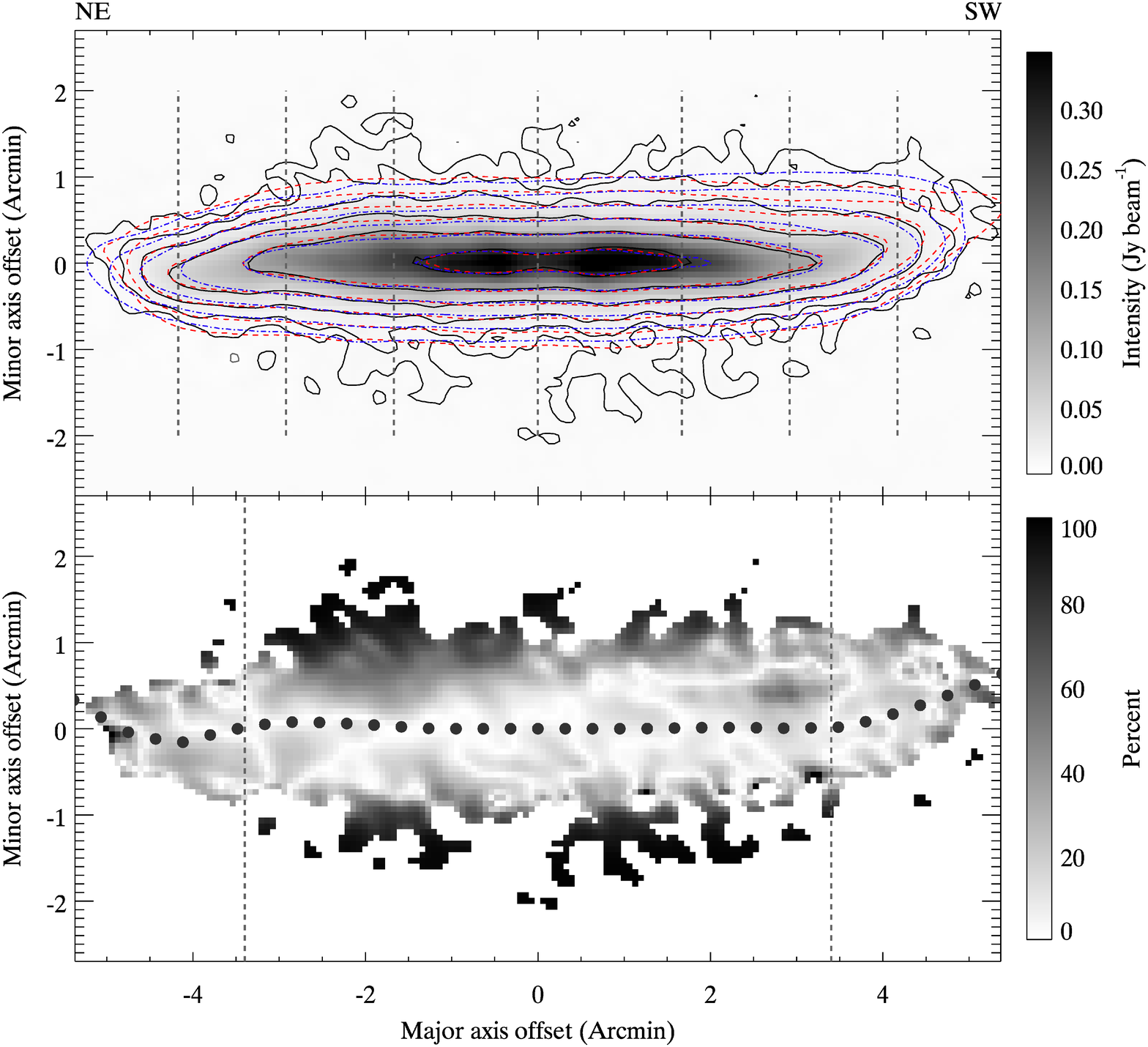} 
   \caption{NGC 5023: Top: Contours of integrated \HI\ (Black Solid Contours, Grey scale), the Base Model (Red Dashed Contours) and the Spiral Arm Model (Blue Dot-Dashed  Contours) at high resolution.  The contour levels are $2.5^n \times 2 \times 10^{19}$\ cm$^{-2}$, $n=1-6$.  Vertical dashed lines indicate the locations of the slices presented in figures \ref{YVmodels} and \ref{YVLocal} Bottom: Residual map of the data and the Spiral Arm model in percentages of the observed flux. Vertical dashed lines indicate D$_{25}$. The dark grey symbols in the lower panel show the run of the PA in the Base Model.}
   \label{mom0Model}
\end{figure*}

For the modeling of the \HI\ we start with first estimates based on values found in the literature (PA, inclination, D$_{25}$ etc.) and derive the initial estimate for the rotation curve from a  map of intensity weighted mean velocities through the {\sc gipsy} routine {\sc rotcur}. This means that this initial estimate will underestimate the real rotation curve  due to the edge-on orientation  of NGC 5023 \citep{1979A&A....74...73S}. To obtain initial estimates of the surface brightness as well as a scale height we compare averaged emission profiles in the vertical direction as well as the radial direction (See Figure \ref{lineprofileHI}). It was found that the shape of the surface brightness profile is very well constrained by the {\sc gipsy} routine {\sc radial}.  We use these initial estimates as input for \tiri\  \citep{Joshtirific2007} and let it fit the central coordinates. These central coordinates are then fixed in the fitting process but refitted for the final models to ensure their accuracy. All models have a velocity dispersion (including instrumental effects) of 10 km s$^{-1}$ \citep[][and references therein]{2011ARA&A..49..301V}.  \\
\indent After fitting the central coordinates we proceed to fine-tune the initial parameters, through an iterative process in which fits by eye are alternated with  $\chi^2$ minimization with \tiri , for a set of possible models. The fits by eye are always done by a comparison between the data and the model in several representations such as PV-diagrams, moment maps, channel maps and radial profiles (shown in Figures \ref{lineprofileHI} to \ref{5023XV}).  The final fit was always done by eye and subsequently confirmed by $\chi^2$ minimization. For simplicity  we compare the models and the data in a coordinate system that is rotated by 62$^\circ$ such that the major axis of the central parts of the galaxy is aligned with the x-axis (as is the case in Figures showing moment or channel maps). Once we obtain the best fit  to the data by varying the PA, scale height, surface brightness profile and rotation curve we investigate how the addition of other features, such as flares, line of sight warps and a lag, influence the fit to the data. We always fit the approaching and receding side independently unless noted otherwise. For a more detailed description of the modeling process see \cite{Kamphuis2011,Zschaechner2011}. We will describe the successes and failures of the individual models below.\\
\subsection{Base Model}
\begin{figure*}
   \centering
   \includegraphics[width=18 cm]{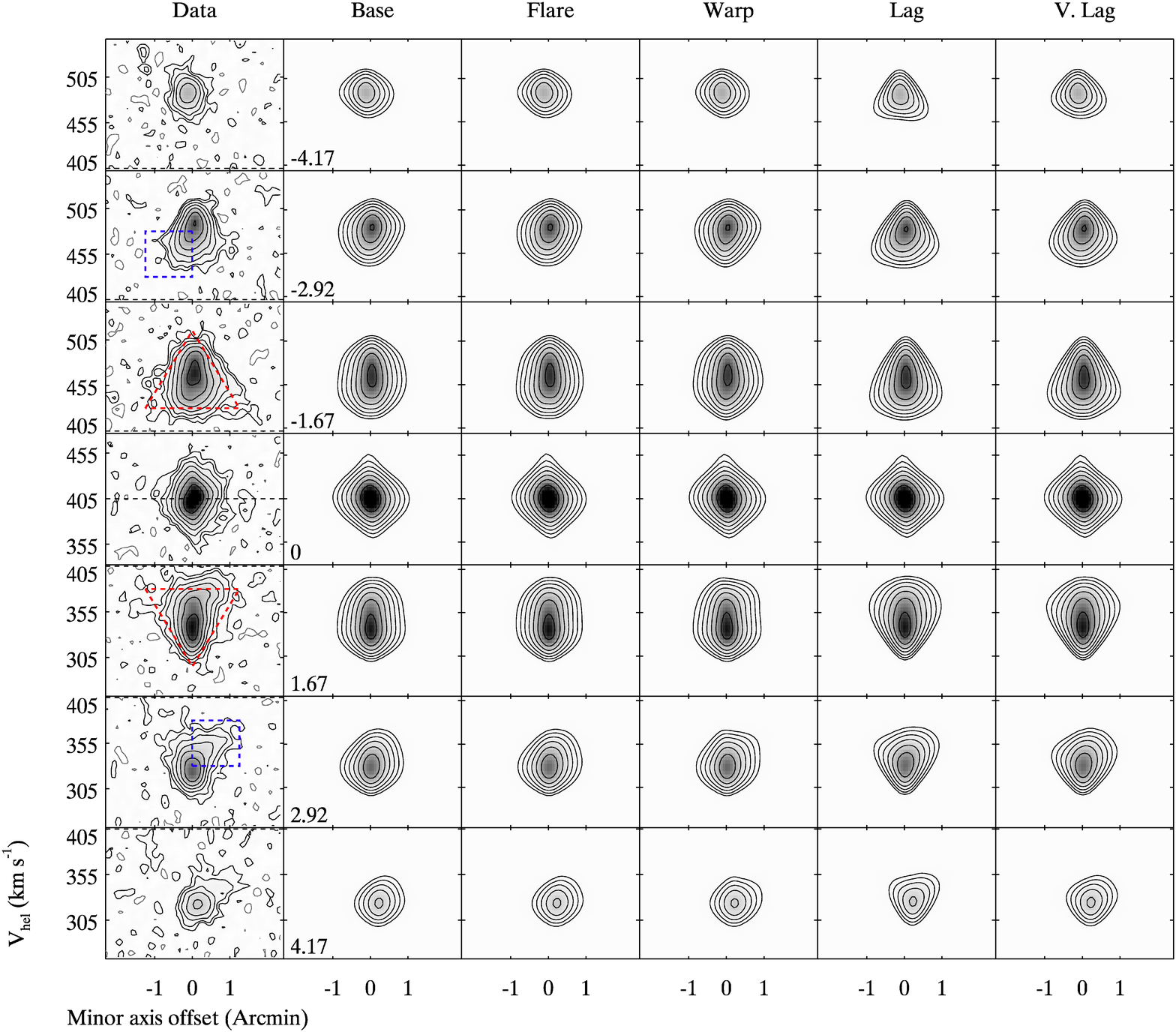} 
   \caption{NGC 5023:  Representative PV-diagrams parallel to the minor
  axis at different major axis offsets (shown in each panel) for the
  data (left) and the various models.  Contour levels are $ 2^n \times 1.5 \times$  the noise level
  of  0.19 mJy beam$^{-1}$, for $n=1-7$ (negative $n=1,2$ (Grey) is also shown).  The models are, from left to right, the Base model, a flaring model, a model with a line-of-sight warp, a model with a vertical gradient and the last column shows a model with a vertical gradient that varies as function of radius. The red and blue dashed lines outline areas of interest which are used in the text to highlight differences between models. The horizontal lines show the systemic velocity.}
   \label{YVmodels}
\end{figure*}

This model is obtained by fine-tuning our initial estimates and will be used as the starting point for all the other models. In this model  we vary only the parameters that need to be varied in order to obtain a minimal satisfactory fit to the data.  In the case of NGC 5023 this means that the surface brightness, the rotational velocities  and the position angle (PA) are required to vary per ring and the approaching and receding sides are fitted independently.  In addition to these parameters that can vary from ring to ring, we also fit the inclination and scale height as a single parameter, i.e. constant for all rings. Figure \ref{lineprofileHI} shows the integrated major axis surface brightness profile of NGC 5023 (right panel, solid black line) as well as the profile of the base model (green asterisks) and the {\sc radial} profile (blue dashed line) used as an initial estimate. The vertical profile (left panel) shows a smoothly declining single exponential profile, with a second component at the higher offsets. However, as will be shown later on, this second component is entirely due to the individual features described in Section \ref{N5023ind}. Therefore we forego fitting a double exponential or double Gaussian layer in our models \citep[see also][]{Zschaechner2011}. With the exclusion of the warped area of the disk, the final parameters in this model vary less than 5\% from their initial estimates, implying that the methods used for obtaining these estimates are fairly accurate for flat disks. \\
\indent The major axis profile in Figure \ref{lineprofileHI} shows a mismatch between the data and the model: the model underestimates the flux in the data around a radial offset $\sim$ -100\arcsec. A zeroth moment map  (Figure \ref{mom0Model})  shows that this is mostly due to the two large individual extra-planar features described in $\S$ \ref{N5023ind} (Features 2 and 3) that can never be captured in global axisymmetric models.  We have confirmed this by comparing the central strip (20\arcsec\ wide) of the model to the data and indeed the discrepancy disappears. These features also cause the bumps seen at column densities $\sim10^{22}$ cm$^{-2}$ in the vertical profiles (Fig \ref{lineprofileHI}, left panel grey dashed lines).  That these bumps are caused by the features can be seen from the difference between the grey dashed line, that includes all the data, and the solid black line, where the areas with a final residual, of the data minus the model, higher than 50\% are excluded. The bottom panel of Figure \ref{mom0Model} shows clearly that these areas are solely due to the described features. From Figure \ref{mom0Model} it also becomes clear that for a minimal satisfactory fit we need to vary the PA as the peak of the emission at the outer radii bends away from the mid-plane. This run of  PA is shown as circular dots in the bottom panel of Figure \ref{mom0Model} and shows that, even though the PA of all rings was fitted as a free parameter, the PA angle hardly varies within D$_{25}$ (vertical dashed lines). \\
\indent This model provides us with a best fit surface brightness profile and rotation curve. Such a model fails to fit the data in several aspects. Figure \ref{YVmodels} shows  PV-diagrams parallel to the minor axis at 5 different radial offsets. In these PV-diagrams  the data clearly show the triangular shape, outlined by the red dashed lines, that is indicative of a vertical gradient in the projected rotational velocities. However, in our simple model (2nd column) these triangular shapes are missing. Especially, the opening angle of the contours on the terminal side is much larger in this model. This behavior is also recognized in the channel maps (not shown) furthest from the systemic velocity where the layer in the model is far too thick when compared to the data. This behavior is seen throughout the galaxy and clearly indicates that the model needs a modification that causes lower projected velocities above the plane. Now we explore whether this can be reproduced with structural changes or only with changes to the kinematics.\\
\subsection{Flare}
One way of obtaining lower projected velocities above the plane could be by including a flare in the model. A flare can be seen as a central depression in the radial gas distribution which grows as one moves away from the mid plane. This way, as one moves away from the mid plane, less and less gas is detected on the line of nodes thus making it appear as if the rotational velocities are declining. In our modeling a flare is simulated by increasing the scale height of the rings at larger radii. To construct a flaring model we refit the scale heights of all rings using once more a mixture of fits by eye and \tiri. To ensure that we obtain the optimal flare model we allow the other parameters to rescale as a whole. This resulted in a minimal change, compared to the base model, in the inclination ($\Delta {\rm inc} =0.02^{\circ}$).\\
\indent Looking at the third column in Figure \ref{YVmodels}, which shows the flaring model, we see immediately that the flare does not significantly improve the match between these models and the data. The flaring model has the same problems as the base model. As a flare would mean another set of free parameters and the fit does not significantly improve compared to the base model we do not include a flare in our final modeling.  
\subsection{Line-of-sight Warp}
A second way to lower the projected line of sight velocities is to include a warp component along the line-of-sight. We can already see in the zeroth moment map (see Fig. \ref{mom0Model}) that the outer radii of the disk are warped. This warp might be purely restricted to the plane of the sky, however it is more likely that it is also partially along the line of sight. Such a warp can be included in the models by letting the inclination vary as a function of radius. Once more this would place gas at the outer radii above the plane  thus leaving the line of nodes empty. The PV-diagrams of this model are shown in the 4th column of Figure \ref{YVmodels}. In this case the allowed rescaling of the other parameters did not lead to a change in these parameters. Like the flaring model, this model does not significantly improve the fit to the data. Most importantly the triangular shape is still not matched in any of these PV-diagrams. \\
\indent We also tried more  extreme versions of a line-of-sight warp and flaring, i.e. ones that were not the best fit to the zeroth moment map or radial/vertical intensity profiles, but none of these improve the shapes seen in the PV-diagrams enough to warrant the degradation of the fit to the density distribution. 
\subsection{Lag}\label{Lag}
The last global option to alter the velocities of the gas above the plane is not to rely on projection effects but to actually lower the rotational velocities as a function of distance above the plane.  This model is shown as the second to last column of Figure \ref{YVmodels}. It can be quite easily seen that of the models discussed so far this is the only model that actually shows the same overall shapes as the data in these PV-diagrams.  The model shown in the figure does not include a flare or a line-of-sight warp thus maximizing the vertical gradient required to match the data. This model was chosen as it has the fewest free-parameters and neither the flaring model nor the line-of-sight  warp model, without a lag, significantly improve the overall fit to the data. However, in order to check quantitatively that the flaring model and line-of-sight  warp model do not affect the determination of the vertical gradient in rotational speeds, a lag was fitted to all three models. \\
 \begin{table}
    \centering
    \begin{tabular}{@{} lcc @{}} 
       \hline
       \hline
       Model & {\sc TiRiFiC} Lag  & Measured Slope\\
       High resolution   	&km s$^{-1}$ kpc$^{-1}$& km s$^{-1}$ kpc$^{-1}$\\
       \hline
Base Model& -14.9 $\pm$ 2.7     & -17.9 $\pm$   5.9\\
 Flare& -15.3 $\pm$ 3.7       & -16.3 $\pm$  6.5\\
 Line of Sight Warp&-14.9 $\pm$ 3.2     & -17.4 $\pm$   6.1\\
 Spiral Arms& -9.4 $\pm$  3.8      & -14.7  $\pm$   8.2\\ 
 Features     & -5.4 $\pm$ 3.2& -10.9 $\pm$  8.6\\
           \hline
        Low resolution & &\\
           \hline
           Base Model&  -11.6 $\pm$ 6.6    & -19.4 $\pm$   3.9\\
 Flare& -13.3 $\pm$ 6.2    & -18.2 $\pm$   4.2\\
 Line of Sight Warp&-12.6 $\pm$ 5.4 & -19.1 $\pm$   4.0\\
 Spiral Arms&  -7.9 $\pm$  4.9 & -15.4  $\pm$   5.2\\
 Features     &  -5.3 $\pm$  5.3 & -8.8 $\pm$   5.1\\
\hline
    \end{tabular}
    \caption{NGC 5023: Vertical gradients that are required for matching the models to the data for both the high resolution data (FWHM = 19\arcsec\ $\times$ 13\arcsec) as well as the low resolution data (FWHM = 36\arcsec\ $\times$ 33\arcsec). Left Column: Name of the model. Middle Column: Vertical Gradient as determined from fitting with {\sc TiRiFiC}. Right Column: Vertical gradient as determined from measuring the slopes of normalized PV-diagrams parallel to the minor axis. }
    \label{tab:lag}
 \end{table}
\indent In order to obtain a objective estimate of the required vertical gradient in these models we have fitted in two different methods. Once by \tiri\ and once by measuring it directly in the model and the data.  The estimate with \tiri\ was done by approaching the lag with initial values on both sides of the expected value. As \tiri\  is used in the Golden Section mode, which fits local minima  \citep[See][]{Joshtirific2007}, the final value of the fit can differ depending on the initial input value. Therefore the lag, and its error,  was determined as the average of two {\sc TiRiFiC} fits; one in which the lag was fitted starting from 0 km s$^{-1}$ kpc$^{-1}$,  and one where the starting point was -22  km s$^{-1}$ kpc$^{-1}$. These lags are presented in Table \ref{tab:lag}. This table also provides the second method where the vertical gradient is measured in normalized PV-diagrams along the minor axis (See \cite{Kamphuis2007a,Kamphuis2011}). In such diagrams, every line profile is normalized by its maximum, and then the slope of maxima as function of height above the plane is measured in the data and subsequently compared to the final model with a range of vertical gradients (e.g. 0-30 km s$^{-1}$ kpc$^{-1}$). The vertical gradient is  found as the gradient in the model where the measured slopes compare best  to the slopes in the data. For NGC 5023 this slope was measured in the vertical range from $\pm$ 10-70\arcsec\  in strips 20\arcsec\ (30\arcsec\ for the low resolution data) wide over a radial range of $\pm$ 50-150\arcsec.  We see that the two methods are formally in good agreement but that measuring the slopes consistently gives higher values. Tests with measurements over different areas show that in general increasing the maximum height of the measurement range lowers the measured vertical gradient by roughly $\sim$ 0.2  km s$^{-1}$ kpc $^{-1}$ arcsec$^{-1}$.  This means that this value is sensitive to the area we choose. Therefore we will rely on the lags determined by {\sc TiRiFiC} for the rest of our discussion as it fits the whole cube while accounting for the noise in the data.\\  
\indent From Table \ref{tab:lag} it becomes clear that, if all we model are axisymmetric structures, models of NGC 5023 need a vertical a vertical gradient of $\sim$-15 $\pm$ 3  km s$^{-1}$ kpc $^{-1}$ in its rotational velocities.  It also confirms our previous conclusion that  the Line-of-Sight Warp and the Flaring models do very little to improve the fit to  the PV-diagrams shown in Figure \ref{YVmodels}. As these models do not significantly improve the fit we conclude at this point that the Base model with a vertical gradient of  $\sim$ -15 $\pm$ 3  km s$^{-1}$ kpc $^{-1}$ is our superior approach in the ``classical'' tilted ring analysis. However, there are still many discrepancies between the data and the model that resembles the data best.\\
\indent  As a last resort in our global models we now try varying the lag in the model from ring to ring. This model is shown in the last column of Figure \ref{YVmodels}. Such a variation has been found in  other galaxies \citep{Oosterloo2007,Zschaechner2011,Zschaechner2012} and if present in all galaxies it could provide us with important clues about the origin of the lagging gas. Additionally, an inspection of Figure \ref{YVmodels} clearly shows that  the lag in our lag model is overestimated at the outer radii (i.e. the bottom and top row in Figure \ref{YVmodels}). In order to investigate such a radial variation we take two separate approaches. Firstly we take our final models with a constant vertical gradient and let \tiri\ fit the gradient in each individual ring. This resulted in the rings having some very different gradients. However, there was no clear radial dependence visible over all rings and the new models were no significant improvement over the models with a constant gradient for all rings. Secondly, we reduced the lag at the outer radii fitting the models by eye, consistent with previous findings of a radially varying lag \citep{Zschaechner2011,Zschaechner2012}. This second method resulted in a lag that declines from -14.9 km s$^{-1}$ kpc$^{-1}$ at 133\arcsec (4.3 kpc) to 0 km s$^{-1}$ kpc$^{-1}$ at 304\arcsec (9.7 kpc). This model improves the fit at the outer radii significantly and is considered at this point the best approach in describing the data. \\
\indent With our ``classical'' analysis, where all variations of the disk extend over half a cylinder, we have now found that the most likely explanation for the kinematics above the plane is a radially varying vertical gradient in the rotational velocities of the gas. Even though the vertical gradient declines to 0  km s$^{-1}$ kpc$^{-1}$ at the outer radii, the PV-diagrams parallel to the minor axis (Figures \ref{YVmodels} and \ref{YVLocal}) show that the opening angle of triangles is now too small, i.e. the lag appears too high. Additionally, the PV-diagrams show that at major axis offsets of 1.7\arcmin, 2.9\arcmin\ and 4.2\arcmin\ there are extensions at mid- to high-altitude (vertical offsets between 0.4\arcmin and 1.3\arcmin) which are not reproduced by the model. These extensions are at low projected rotational velocities (V$_{\rm hel}$ 25 to 75 km s$^{-1}$ away from V$_{\rm sys}$) and are outlined by the blue dashed boxes in Figure \ref{YVmodels}. Similar extensions appear in the PV-diagrams at major axis offsets of -1.7\arcmin\ and -2.9\arcmin, albeit to a lesser degree, and on the other side of the disk (at V$_{\rm hel}$ 0  to  60 km$^{-1}$ away from V$_{\rm sys}$ and vertical offsets between  -0.4\arcmin and -1.3\arcmin).  Clearly they are extended structures in, or right above, the disk that are currently not reproduced by our global models. \\
\subsection{Non-symmetric Modeling}\label{features}
The current version of {\sc TiRiFiC} allows us to add partial (i.e. of restricted azimuth range) disks to our models. This means that we can now construct features in our models which cover only a small part of the total disk and that have parameters that differ from the underlying axisymmetric disk.  This enables us to investigate the effect of localized features on our global analysis. The data already suggest that the extra-planar gas in NGC 5023 is far from cylindrically symmetric and therefore we think this galaxy is an excellent opportunity to test this next step in tilted ring modeling. As discussed before, when the density distribution of the extra-planar gas differs significantly from that in the disk, this can mimic a vertical gradient due to gas missing at the line of nodes.\\
\indent As a proof of concept we try to model the individual features first by adding a partial disk to each side (i.e. approaching and receding) of the final models of the previous sections. These disks are limited to a maximum azimuthal extent of 90$^{\circ}$, i.e. a quarter disk, and are restricted to follow the rotation curve and PA  of the final model of the previous section. However, we refit the radial surface brightness profile and the scale height of the model. We also fit  the scale height, surface brightness profile, azimuthal extent and azimuthal center, i.e. the PA of the feature in the face on view, of the new disks. As it is likely that small variations in inclination can now severely affect our results the inclination is treated in a special manner. The rings with radii larger than 175\arcsec\ are fitted independently from each other whereas the rings at smaller radii are fitted with a single value. In this we always require the inclination for the partial disk  to match that of the full disk at any given radius, i.e. the partial disk is oriented in the same way as the full disk. Here, the fitting process is once more a mixture of $\chi^2$ minimization with \tiri\ and fits by eye. The approaching and receding sides are fitted independently.\\ 
\begin{figure*}
   \centering
   \includegraphics[width=17 cm]{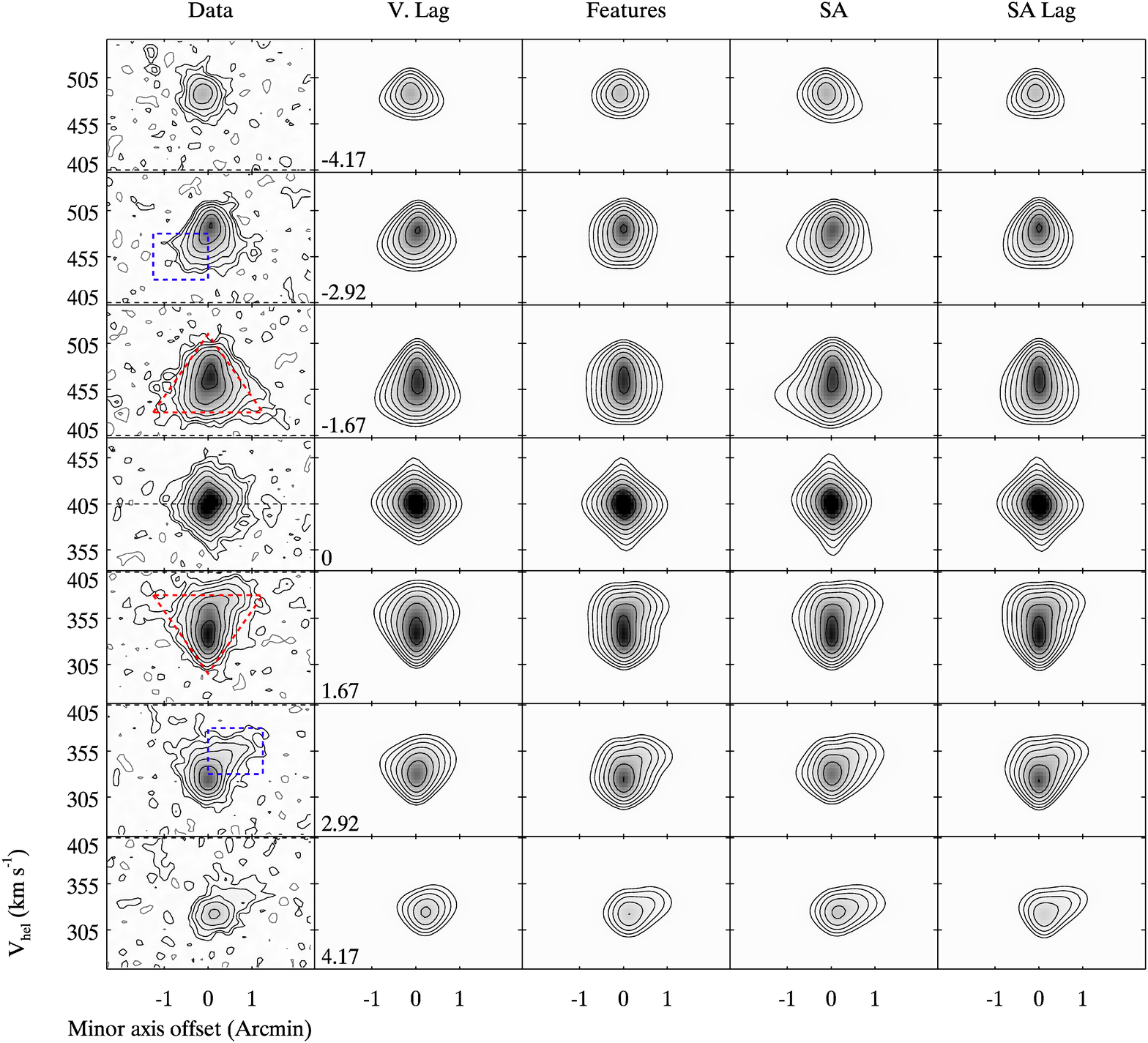} 
   \caption{NGC 5023: As Figure \ref{YVmodels}. The models are, from left to right, the  ``classical" model with a radially varying lag, a model with partial disks added to it, a model with spiral arms added to the disk and the same model with a vertical gradient. The Features model is described in $\S$ \ref{features} and the Spiral Arms models in $\S$ \ref{SA}.}
   \label{YVLocal}
\end{figure*}
\begin{figure*}
   \centering
   \includegraphics[width=17 cm]{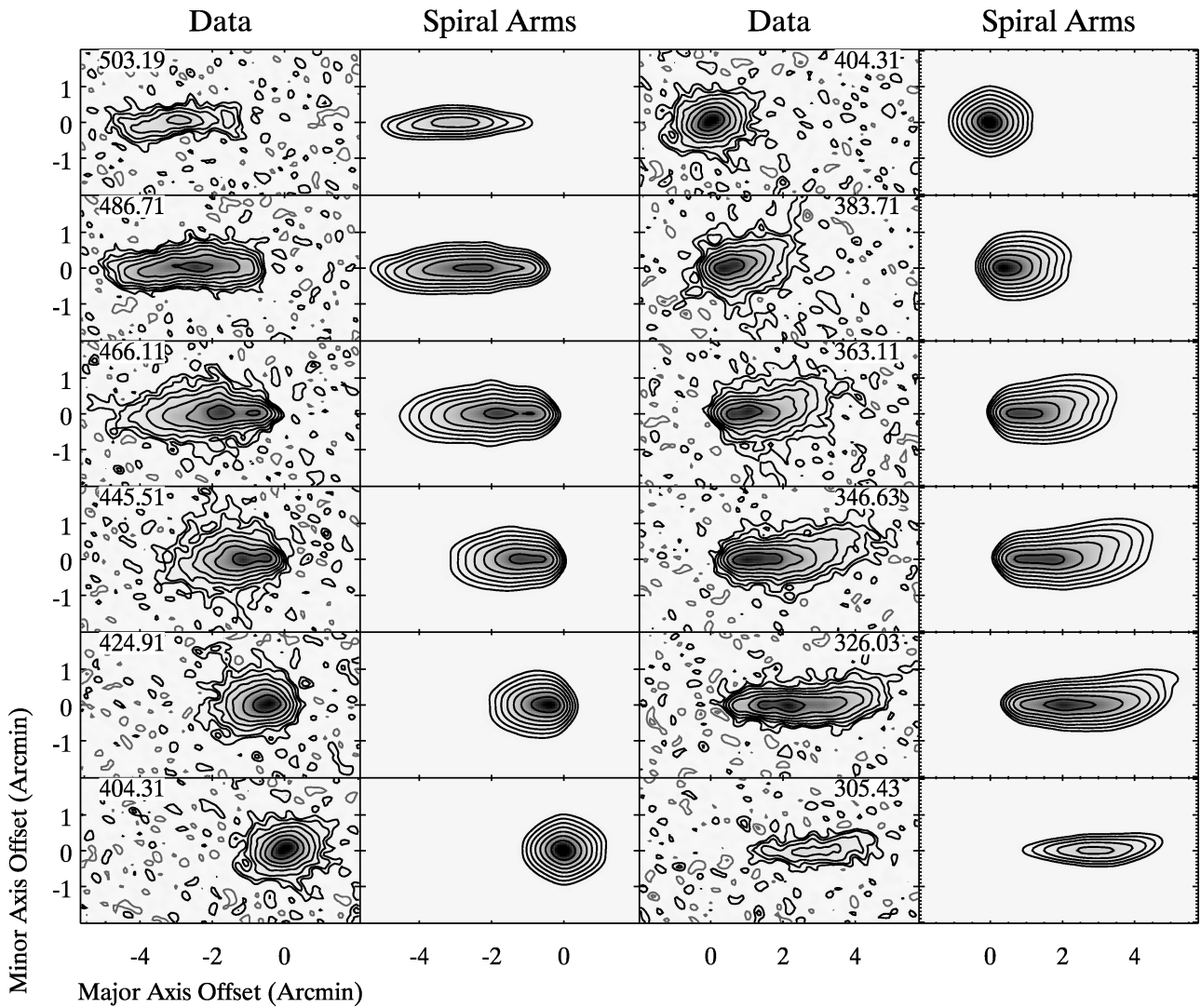} 
   \caption{NGC 5023: Representative channel maps in the rotated frame for the data
  and the best model (``Spiral Arms''). Contour levels are the same as in Figure \ref{YVmodels}.}
   \label{5023Chans}
\end{figure*}
\begin{figure*}
   \centering
   \includegraphics[width=17 cm]{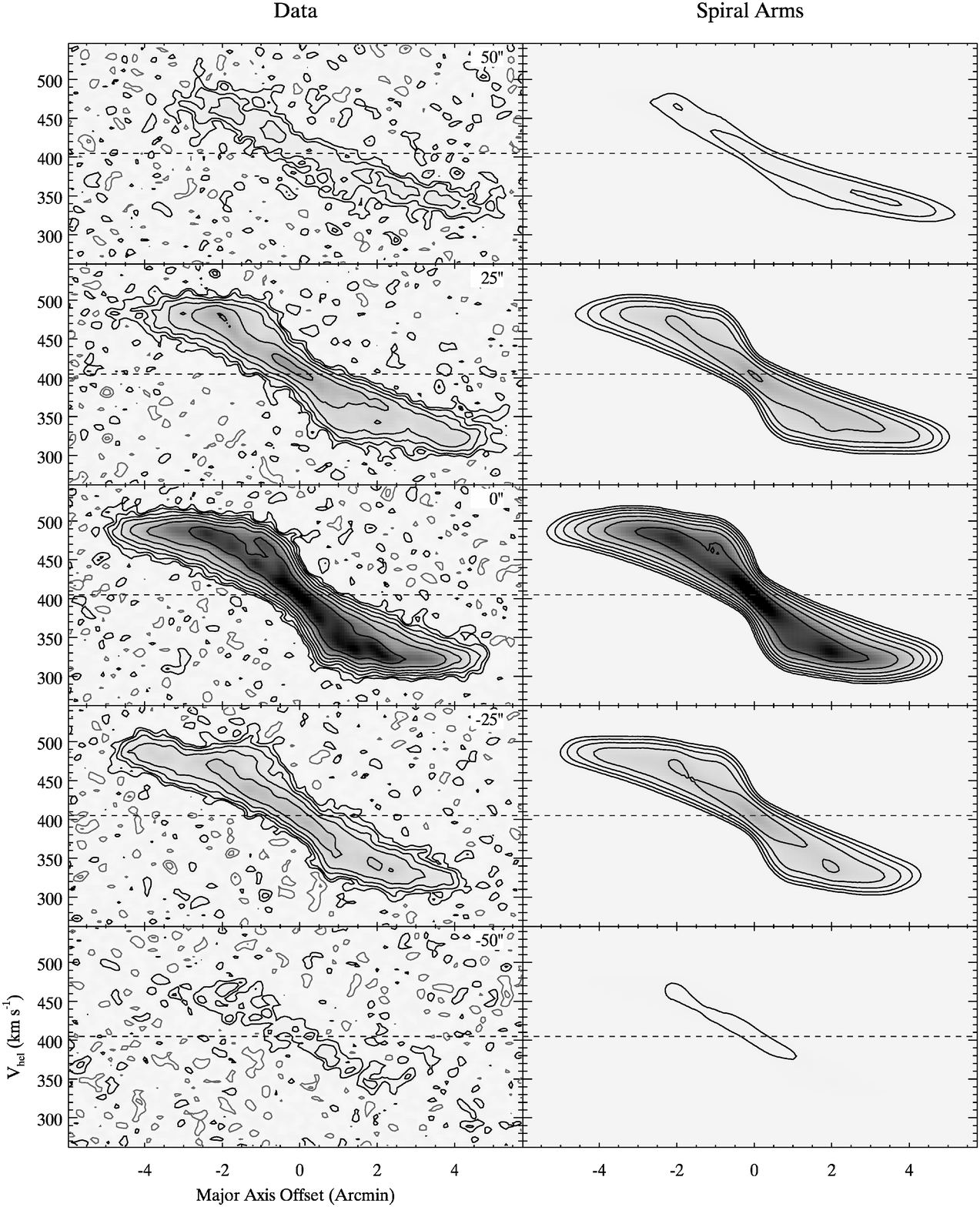} 
   \caption{NGC 5023: Representative PV-diagrams parallel to the major axis at different minor axis offsets (shown
in each panel; positive offsets are to the north in the rotated frame) for
  the data (left) and the best model (right).  Contour levels and dashed lines are the
  same as in Figure \ref{YVmodels}.}
   \label{5023XV}
\end{figure*}
\indent Figure \ref{YVLocal} shows that adding partial disks to the base model  causes a significant improvement immediately. The peculiar extended features seen in Figure \ref{YVmodels} at mid altitude and low rotational velocities are now clearly reproduced on the South Western side of the disk, i.e. positive radial offsets. However, the triangular shapes in the PV-diagrams parallel to the minor axis that are normally associated with a lag are still not defined very well. When we fit a lag for this model in the same way as for the global models we find that the need for a lag is significantly lowered. Even though a best fit lag is still non-zero,  a model without a lag cannot be excluded anymore (See Table \ref{tab:lag}).  However, when we view this model from a face-on perspective, it becomes clear that it is very unphysical. The left hand side of Figure \ref {Face-Onview} shows this face on view, which is created by taking the absolute value of the model's inclination in each ring and subtracting 90$^{\circ}$. From this viewing angle  it is clear that  certain parts of the disk are much brighter. By giving these parts a different scale height (z$_{\rm disk}$ =   8.9\arcsec\, 7.6\arcsec\ (0.28, 0.24 kpc) vs. z$_{\rm feature}$ = 26.1\arcsec, 15.7\arcsec\ (0.84, 0.50 kpc) for the negative and positive offsets, respectively) and adding a line of sight warp the density distribution above the plane is significantly different from that in the plane of the galaxy. Figure \ref {Face-Onview} also shows that the fitting process has pushed the partial disks as far as possible from the line of nodes, therefore minimizing the need   for a vertical gradient. However,  as is easily seen from Figure \ref{Face-Onview} this is not a very realistic distribution of the gas.\\ 
\begin{figure*}
   \centering
   \includegraphics[width=18 cm]{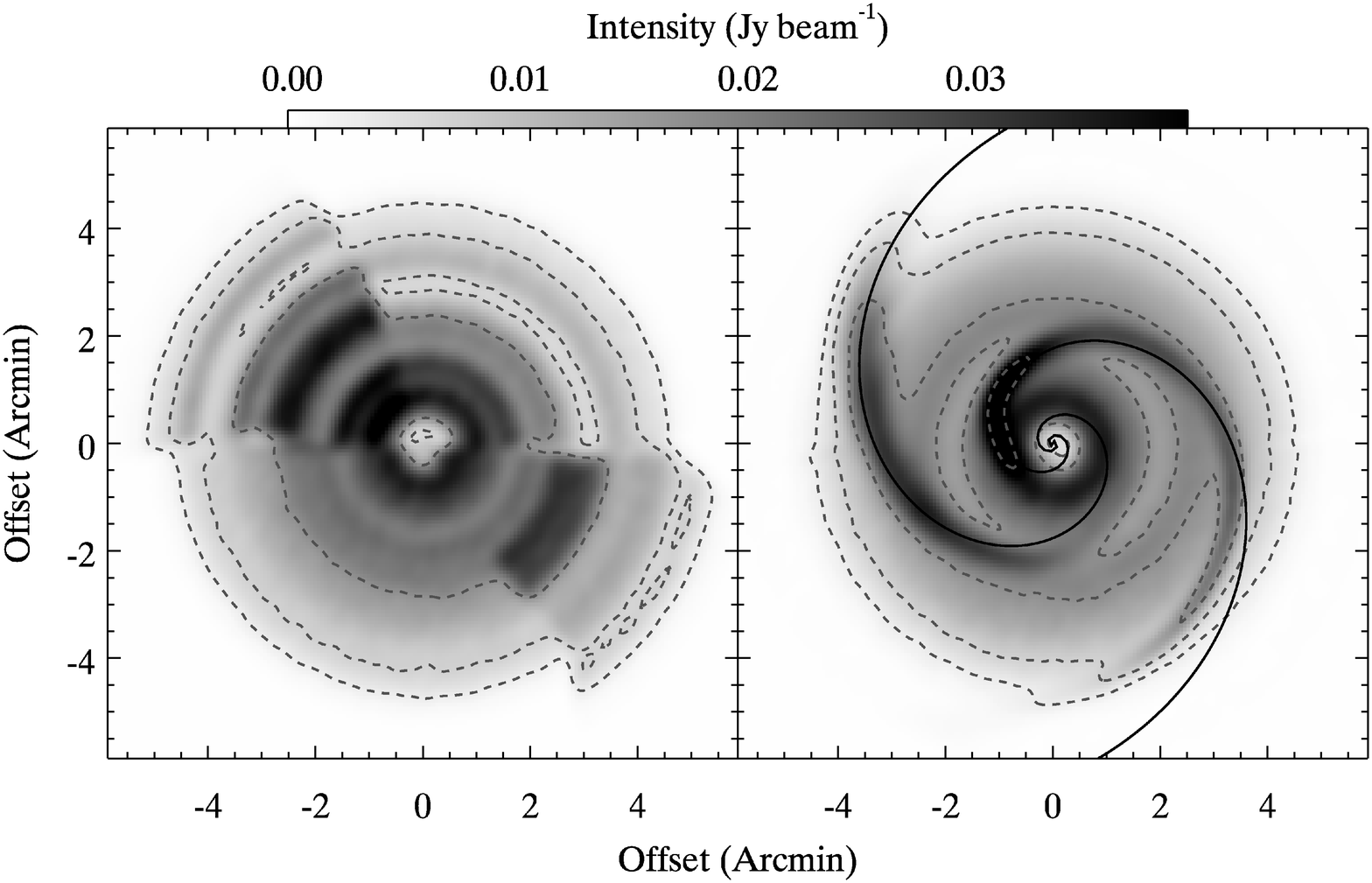} 
   \caption{NGC 5023: Face-on zeroth moment map of the model with partial disks added (Left) and spiral arms (Right). The contour levels are $2.5^n \times 2 \times 10^{19}$\ cm$^{-2}$, $n=1-4$. The black lines in the right hand side map represent a logarithmic spiral arms with a pitch angle of 22$^{\circ}$ following the orientation of the disk. The normal line-of-sight is along the x-axis with the observer on the right hand side.}
   \label{Face-Onview}
\end{figure*}
\subsection{Spiral Arms}\label{SA}
\begin{figure}
   \centering
   \includegraphics[width=8 cm]{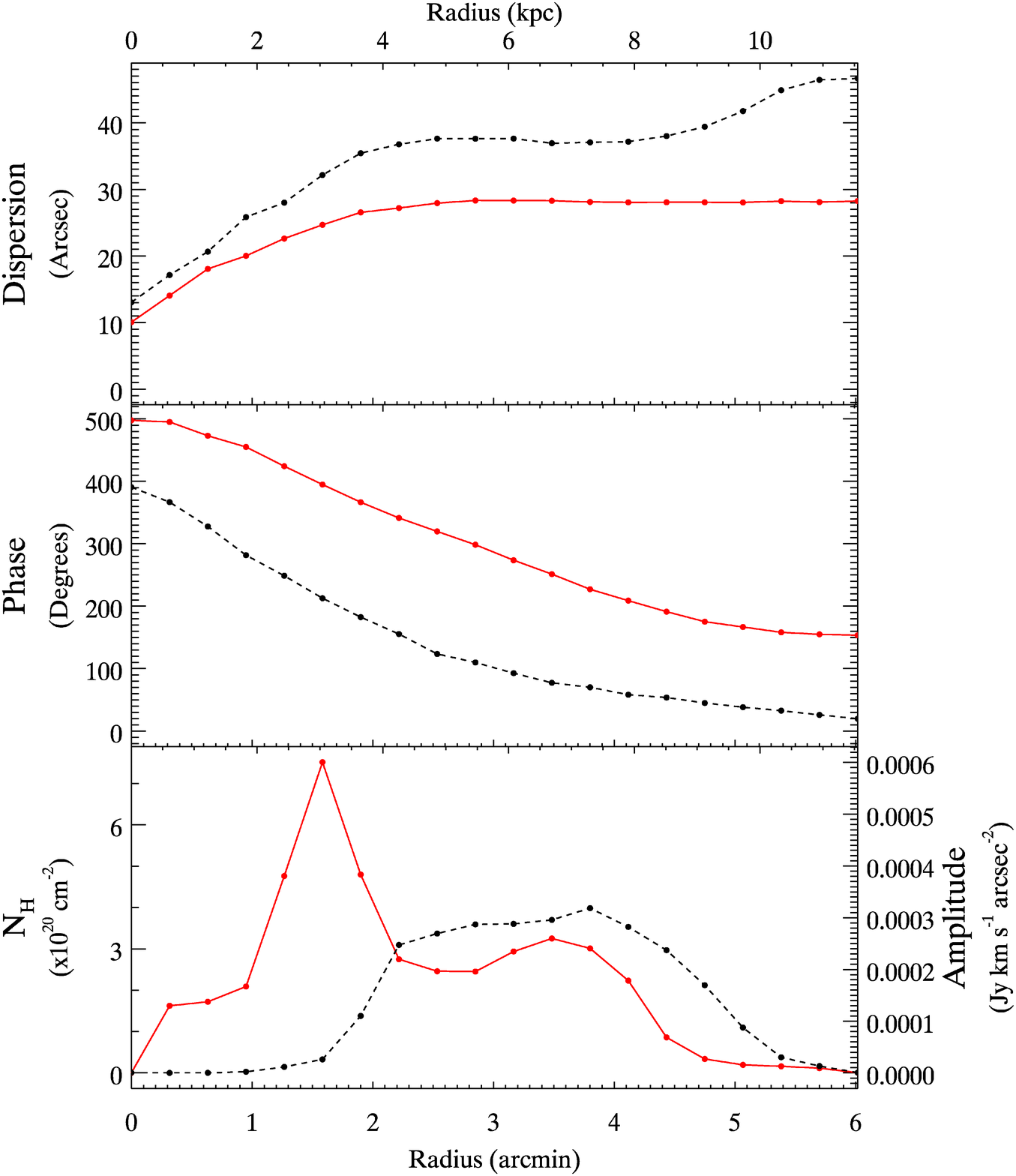} 
   \caption{NGC 5023: Parameters of the arms in the best fit model with Spiral Arms, with from top to bottom: the dispersion of the overdensity/arm, i.e the width in each ring, the phase, i.e. the central position of the overdensity in each arm and the amplitude of the overdensity/spiral arm. The black dashed line and the red solid line show the parameters for the two arms.}
   \label{Parametersarm}
\end{figure}

Since the appearance of the model in the left panel of Figure \ref{Face-Onview} does not resemble typical real galaxies, we try to reproduce the effect while constraining the additional disks to resemble logarithmic spiral arms. We apply the same constraints as in the case of the partial disks but with the additional restrictions that the width and phase angle of the additions follow a smooth distribution that resembles two independent spiral arms. This smooth distribution is achieved through manual regularization. The initial best guess phase angles are calculated such that two perfect identical logarithmic spiral arms with a pitch angle of 20$^{\circ}$ are added to the model. This is somewhat below the typical value for galaxies with rotational velocities similar to NGC 5023 \citep{1981AJ.....86.1847K}.  During the fitting the only constraint on this phase angle is that the changes from one ring to the other are smooth (See Figure \ref{Parametersarm}), i.e. we do not enforce logarithmic spiral arms. The phase went through several, both manual and \tiri, iterations before arriving at this final solution. Another difference in the fitting process is that we refit the PA in a similar manner to the inclination. This is because spiral arms can imitate warping behavior when the inclination deviates from exactly edge-on \citep{1978ApJ...222..815B}. For this reason we may already be interpreting some spiral structure as PA variations in e.g. the Base Model.\\
\begin{figure}
   \centering
   \includegraphics[width=8 cm]{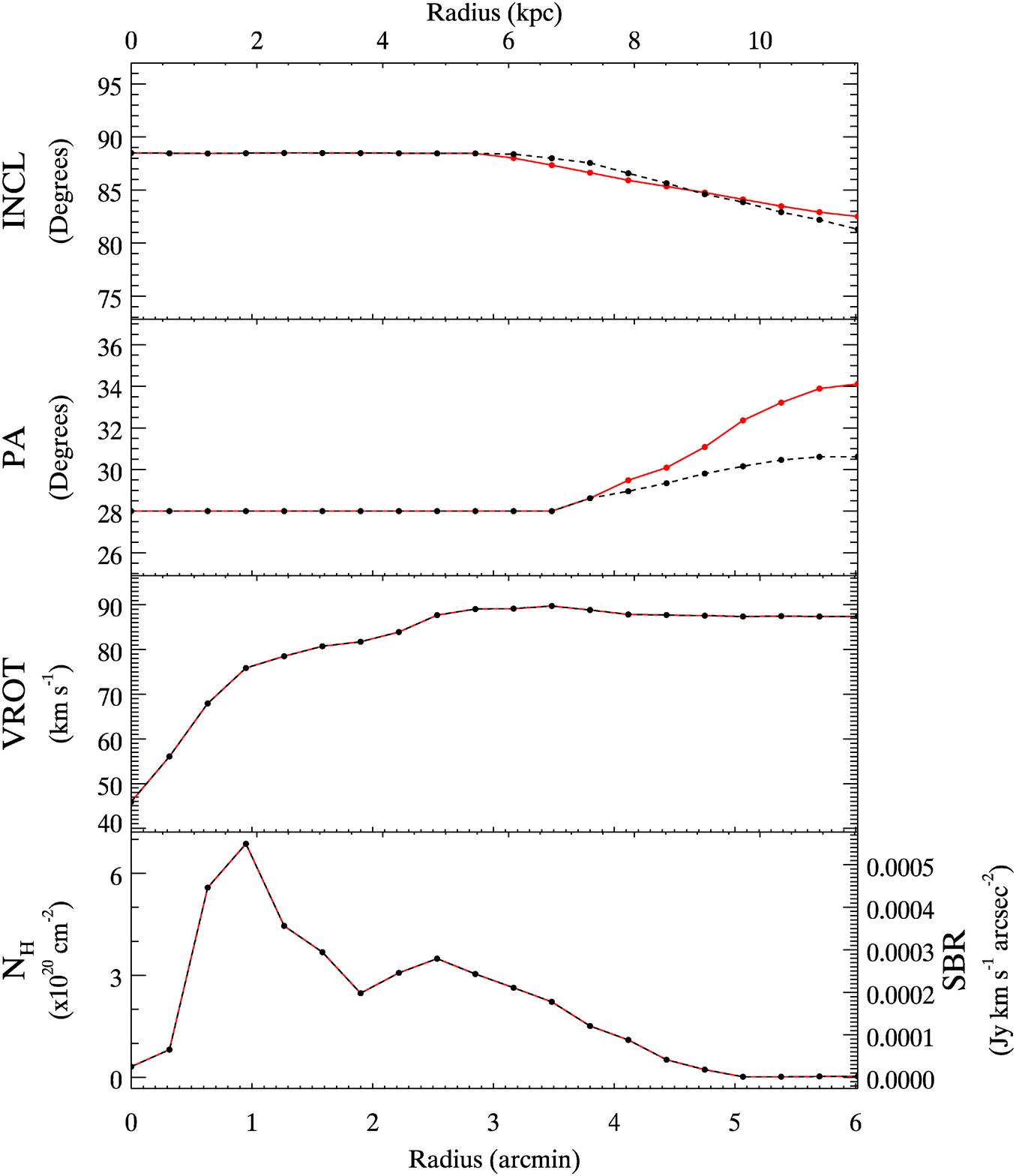} 
   \caption{NGC 5023: Parameters of the disk in the best fit model with Spiral Arms. The black dashed line is for the approaching (SW) side, this line is missing for the bottom two panels as the disks are the equal. The red solid line for the receding (NE) side.  }
   \label{ParametersSA}
\end{figure}
\indent The PV-diagrams of the best fit spiral arms model also successfully  reproduce the low velocity high altitude features outlined by the blue boxes in Figures \ref{YVmodels} and \ref{YVLocal} and now, when seen face on, we have a much more realistic distribution of the gas (See Figure \ref{Face-Onview}, right panel).  The face on view also shows that the added individual structures still roughly follow the shape of logarithmic spiral arms, albeit the pitch angle has changed to $\sim$22$^{\circ}$ (black lines in Figure \ref{Face-Onview}), which is still low for this type of galaxy  \citep{1981AJ.....86.1847K}. Additionally, the underlying disk is now fully cylindrically symmetric except for PA and inclination (See Figure \ref{ParametersSA}), thus significantly reducing the complexity of the models. However,  comparing the opening angles of the model contours in Figure \ref{YVLocal} to those of the data, it is clear that the model still requires a lag. This is confirmed by fitting the lag with \tiri\ in the same way as for the ``classical'' models (See $\S$ \ref{Lag}). This way we find that this best fit model still requires a vertical gradient of $-9.4\pm3.8$ km s$^{-1}$ kpc$^{-1}$. Additionally, we once more investigate the presence of any radial dependence in the vertical gradient but for these models the improvement is insignificant and therefore we discard such models as they add another set of free parameters. However, we would like to point out once more that such a variation does seem to be present when comparing the models to the data. The prime example of this can be seen in the PV-diagram with a 4.2\arcmin\ offset shown in Figure \ref{YVLocal}.\\
\indent Since, the assumption of spiral arms is physically motivated and the addition significantly improves the fit to the data, we deem the Spiral Arm $+$ lag model to be the best fit to data. It must be noted that in this model it is crucial to have a radial variation in the inclination in the disk, i.e. a line-of-sight warp, to fit the extended low velocity high altitude structures. Also, the best fit scale height of one of the arms  deviates from that of the disk and the other arm (See Figure \ref{Parametersarm}, Table \ref{tabX}). Without these changes, especially  the warp, the improvement compared to the Base Model would not be significant. In order to determine which of these parameters is the dominant effect on the lag we refitted the lag for two models: one with the scale height of the arms fixed to the scale height of the disk and one where the inclination of the disk and arms was kept constant at the central value. When the scale heights were fixed this resulted in a lag very similar to the best fit ( dV/dz=-8.8 $\pm$ 2.5 km s$^{-1}$ kpc$^{-1}$). However, when the line of sight warp was removed  the lag (dV/dz=-14.5 $\pm$ -3.1 km s$^{-1}$ kpc$^{-1}$) was found to be  very similar to the final ``classical'' model. This confirms that in the case of NGC 5023 the lowered lag is due to the presence of the arms in the line of sight warp.  \\
\indent It is difficult to provide errors on the different values of the various components of the model, especially when each ring is allowed to vary independently.  This is because many of the parameters are degenerate at some level. In the individual rings the matter is even worse as often the value of each ring is slightly degenerate with its neighboring rings as well. However, in order to provide some estimate of the error on each parameter we have systematically varied each parameter of the best fit model, while keeping the others fixed, until the model and data clearly deviated. If during this process, when lowering the parameter, a value in a single ring became less than zero it was set to zero. The central best fit parameters and their errors determined this way are presented in Table \ref{tabX}. The stated errors are the average maximum allowable deviation found when increasing and decreasing the parameter's values. Because Table \ref{tabX} provides the central values of all parameters it sometimes appears as if the error is larger than the value, however a quick look at Figures \ref{Parametersarm} and \ref{ParametersSA} shows that this is not the case for the majority of the rings.\\
 \begin{table*}
    \centering
    \begin{tabular}{@{} llll @{}} 
       \hline
       \hline
Parameter & Disk   & Arm 1 & Arm 2  \\
       \hline
Center ($\alpha$ J2000)&13$^{{\rm h}}$  12 $^{{\rm m}}$11.83$^{{\rm s}}$ $\pm$ 4.3$^{{\rm s}} $&& \\
	 \hspace*{0.93cm}($\delta$ J2000)&44$^{\circ}$2\arcmin\ 16.9\arcsec\ $\pm$ 6 \arcsec& &\\	
$V_{\rm sys}$ (km s$^{-1}$)&404.8 $\pm $ 4& &\\
Inclination ($^{\circ}$)&88.5 $\pm$ 1.3& & \\
PA &28.0 $\pm $1& &  \\
$V_{\rm rot}$ (km s$^{-1}$)&87.6 $\pm $ 5& & \\
Scale height (arcsec)&9.1 $\pm$ 3&14.0 $\pm$ 9&9 $\pm$ 6\\
Scale height (kpc)&0.29 $\pm$ 0.10 &0.45 $\pm$ 0.29&0.29 $\pm$ 0.19\\
Mass (\msun)&4.8 $\pm$ 0.26 $\times 10^8$&4.9 $\pm$ 0.7$\times10^7$&4.3 $\pm$ 0.3$\times10^7$ \\
Dispersion (arcsec) & &13 $\pm$ 23&10 $\pm$ 25\\
Phase Center ($^{\circ}$) & &390.8 $\pm$ 7.5& 497.9 $\pm$ 7.5\\
\hline
    \end{tabular}
    \caption{NGC 5023: Parameters of the best fit model and their errors. Except for the mass (total flux) and the rotational velocity (velocity at D$_{\rm{H \textsc{i}}}$) the central values are reported. }
    \label{tabX}
 \end{table*}
\indent The addition of spiral arms introduces a large number of extra parameters to the model. Additionally, we are trying to fit structures that are actually only indicated in the velocity structure of our data and that are normally only visible in face-on galaxies. With 20 rings in each disk of the model  we introduce 122 (Z0 (2), arm amplitude (40), phase center (40), arm dispersion (40))  new free parameters to the model by introducing the spiral arms. Of these, 80 are correlated in the sense that  ring to ring changes should be smooth, therefore large jumps are smoothed in the model.\\
\indent Counting each ring as a free parameter should be considered a simplistic attempt to quantify the number of free parameters in the models. Often user interaction will reduce the freedom of a parameter  such that it should not be considered a truly independent parameter. Additionally the number of rings used for the model can also vary, using smaller ring sizes would result in more rings to vary and thus in more free parameters in our estimation. However, their spacing would be less than a beam and therefore neighbouring rings would be much more correlated than in the current model. Another uncertainty in the number of rings is introduced by the edge of the disk, which is not observed. In the case of NGC 5023 we choose to model the galaxy up to the minimum radius that fully includes the 2$\sigma$ contour in the low resolution cube ($\sim$ 6\arcmin). However, one could argue for more rings, as the edge appears to be undetected, or less rings, as the data become severely affected by the noise.\\
\indent Following the simple approach outlined above, the total number of free parameters for our final model then becomes 206 (Z0 (1), DVRO (1), SBR(20), VROT(20), PA (21), INCL (21)\footnote{The inner ten rings of the disk are fitted as a single value  for the inclination and PA, as described previously in the text}, Spiral Arms). When we compare this to the commonly accepted ``classical''  two disk modeling, we have significantly increased the number of free parameters, as in such a model we would end up with 126 (SBR(40), VROT(40), PA (40), INCL (2), Z0 (2), DVRO (2)) free parameters.  However, the inclusion of common variations such as a radially varying lag, a flare, a free central position, a line of sight warp  or a  2-component vertical disk  would quickly increase the number of free parameters to our spiral arm model. Additionally, we can see in Figure \ref{ParametersSA} that the differences between inclination on both sides of the disk are minimal. Another thing to keep in mind is that the models are still severely oversampled as a conservative estimate give us $\sim$1300 resolution elements (based on an independent area of emission of 250\arcsec$\times$100\arcsec every 3 channels) in the high resolution cube. Therefore we deem the increase of free parameters in the model acceptable, especially as specific features in the data can only be reproduced when such non-cylindrically symmetric overdensities are introduced to the model and  spiral arms are a well documented feature of face-on galaxies.\\
\indent It is remarkable to see how well the shapes of the initial logarithmic spiral arms are maintained (See Figure \ref{Face-Onview}) by the simple condition that the phase angle and width should only vary smoothly, even though their actual pitch angle has changed by almost 2 degrees in the fitting process. Although it must be said that the characteristics of the arms are ill constrained (See Table \ref{tabX}).\\
\indent In general there are three firm conclusions to take away from the modeling:
\begin{enumerate}
\item NGC 5023 requires non-axisymmetric overdensities in its gas distribution. For these overdensities to significantly improve the fit the disk needs to be slightly warped in the line of sight.
\item NGC 5023 requires a vertical gradient in its rotational velocities.
\item In NGC 5023 the value of the vertical gradient is affected by the specifics of the non-cylindrical distribution.
\end{enumerate}

\section{UGC 2082 Models} \label{ModelsU2082}
\begin{figure}
\centering
\includegraphics[width= 8 cm]{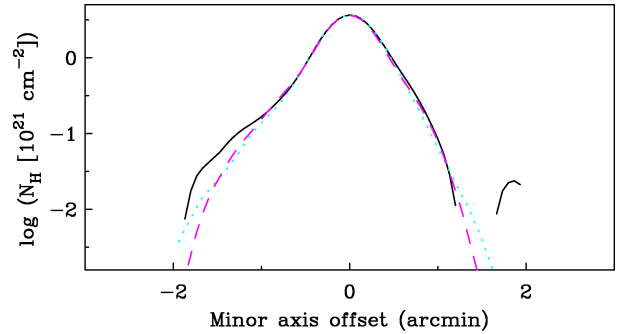}
\caption{UGC 2082: Minor axis profiles averaged over the central 5\arcmin\  along the major
axis for the data (solid black line), base warp model (dotted cyan line),
and best model (dashed magenta line).  Other models (not shown) fit the data about as well but are rejected for other reasons (see text).}
\label{fig1}
\end{figure}

\begin{figure*}
\centering
\includegraphics[width= 12 cm]{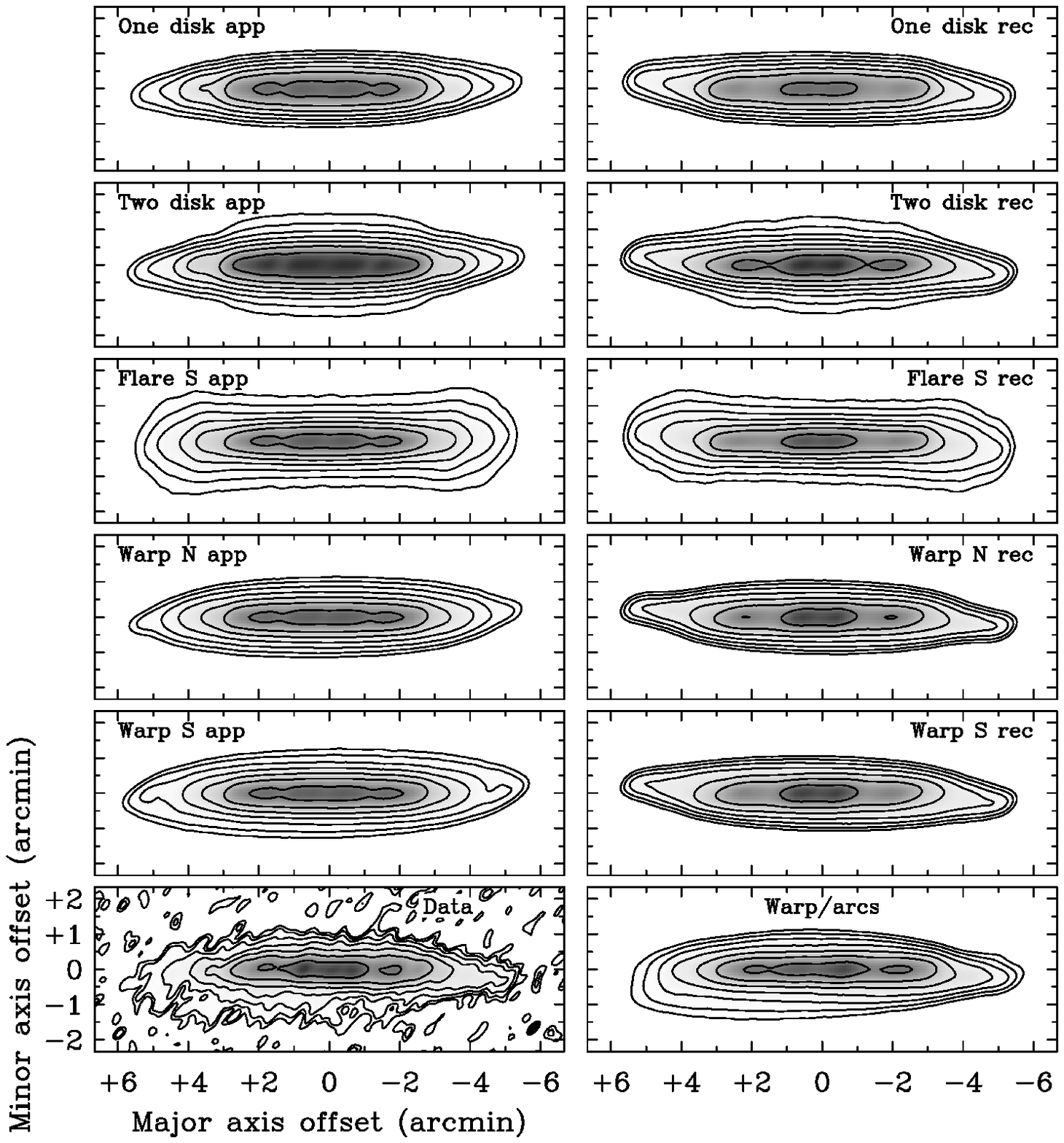}
\caption{UGC 2082: zeroth moment maps for the various models described in the
  text and for the data in the rotated frame where the major axis of
  the unwarped part of the disk is horizontal.  The approaching side
  is at positive major axis offsets.  In this and subsequent figures,
labels ``N'', ``S'', ``app'', and
``rec'' refer to models that are optimised for the North, South,
approaching or receding sides, respectively.  ``Warp'' refers to the
line-of-sight warp in those models.  Contour levels are $2^n \times
  5.1 \times 10^{19}$\ cm$^{-2}$, $n=0-6$.}
\label{fig2}
\end{figure*}

\begin{figure*}
\centering
\includegraphics[width= 13.5 cm ,angle=-90]{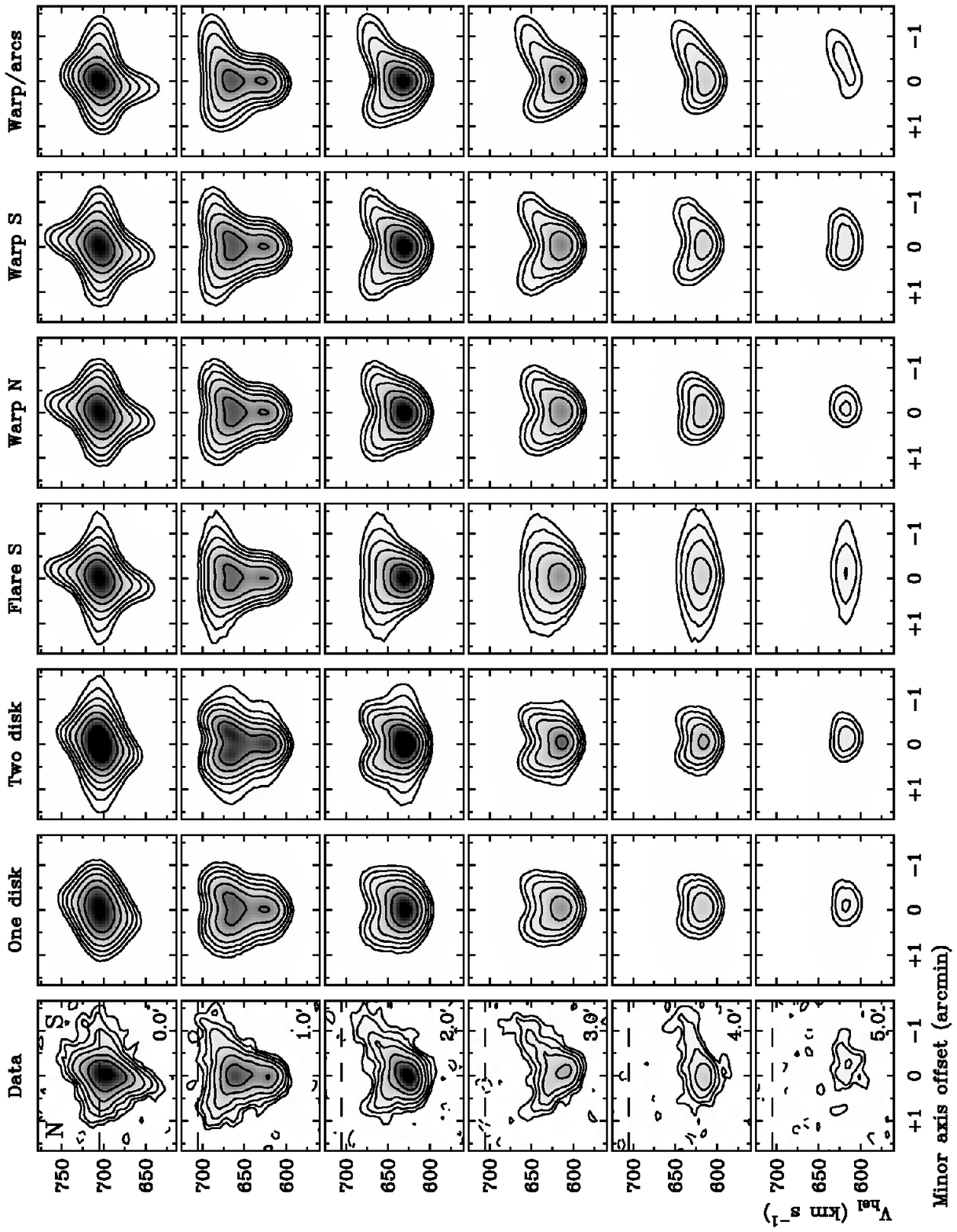}
\caption{UGC 2082: Representative PV-diagrams parallel to the minor
  axis at different major axis offsets (shown in each panel) for the
  data (left) and the various models for the approaching side (as well
  as the minor axis).  Dashed lines in the data panels indicate the systemic velocity}. Contour levels are $2^n \times$ the noise level
  of  0.21 mJy beam$^{-1}$, $n=1-7$ \bf{(negative $n=1$ also shown for the data.}
\label{fig5a}
\end{figure*}

\begin{figure*}
\centering
\includegraphics[width= 11.5 cm,angle=-90]{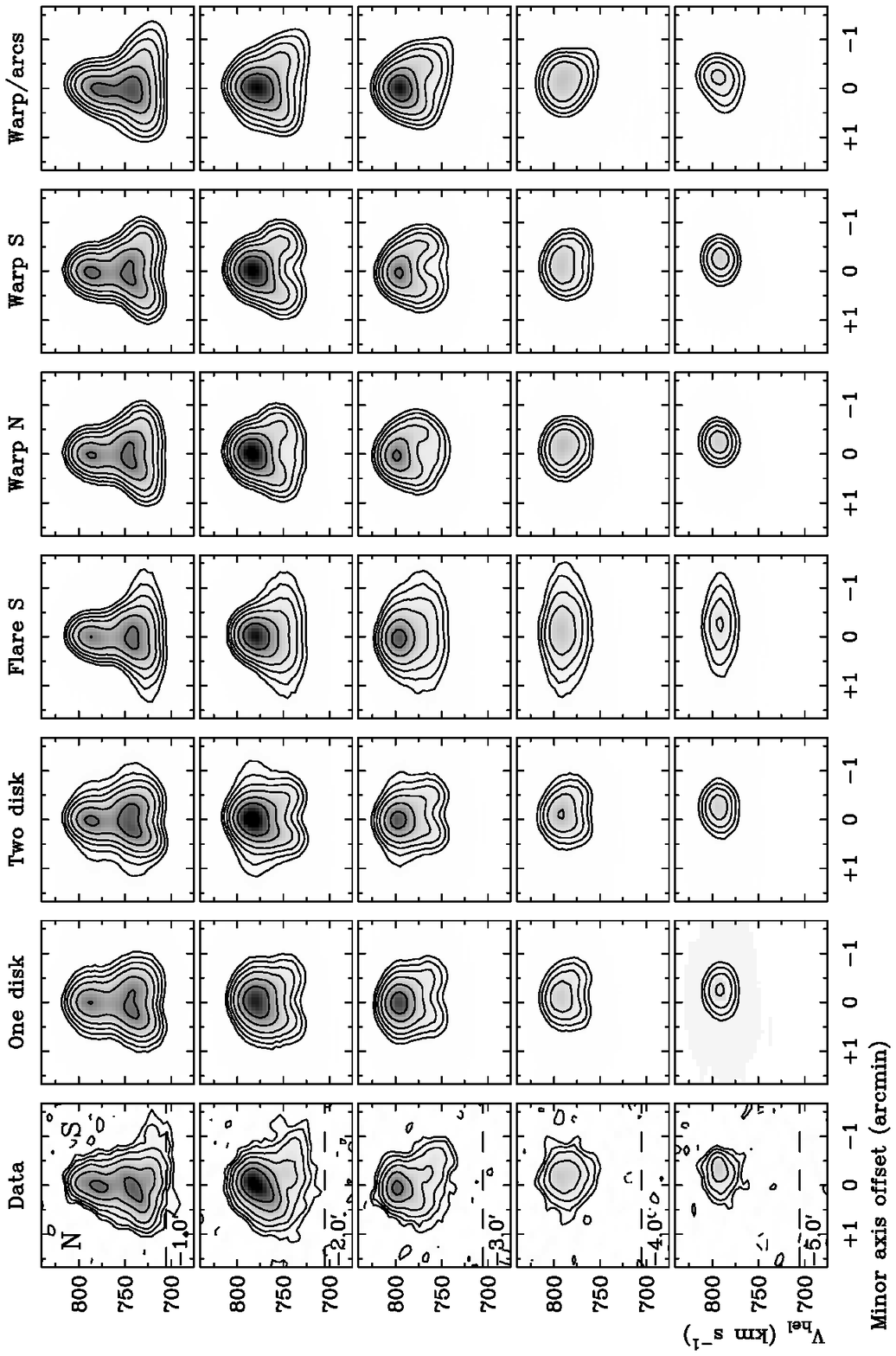}
\caption{UGC 2082: As Figure \ref{fig5a} but  for the receding side.}
\label{fig5b}
\end{figure*}

\begin{figure*}
\centering
\includegraphics[width= 16 cm]{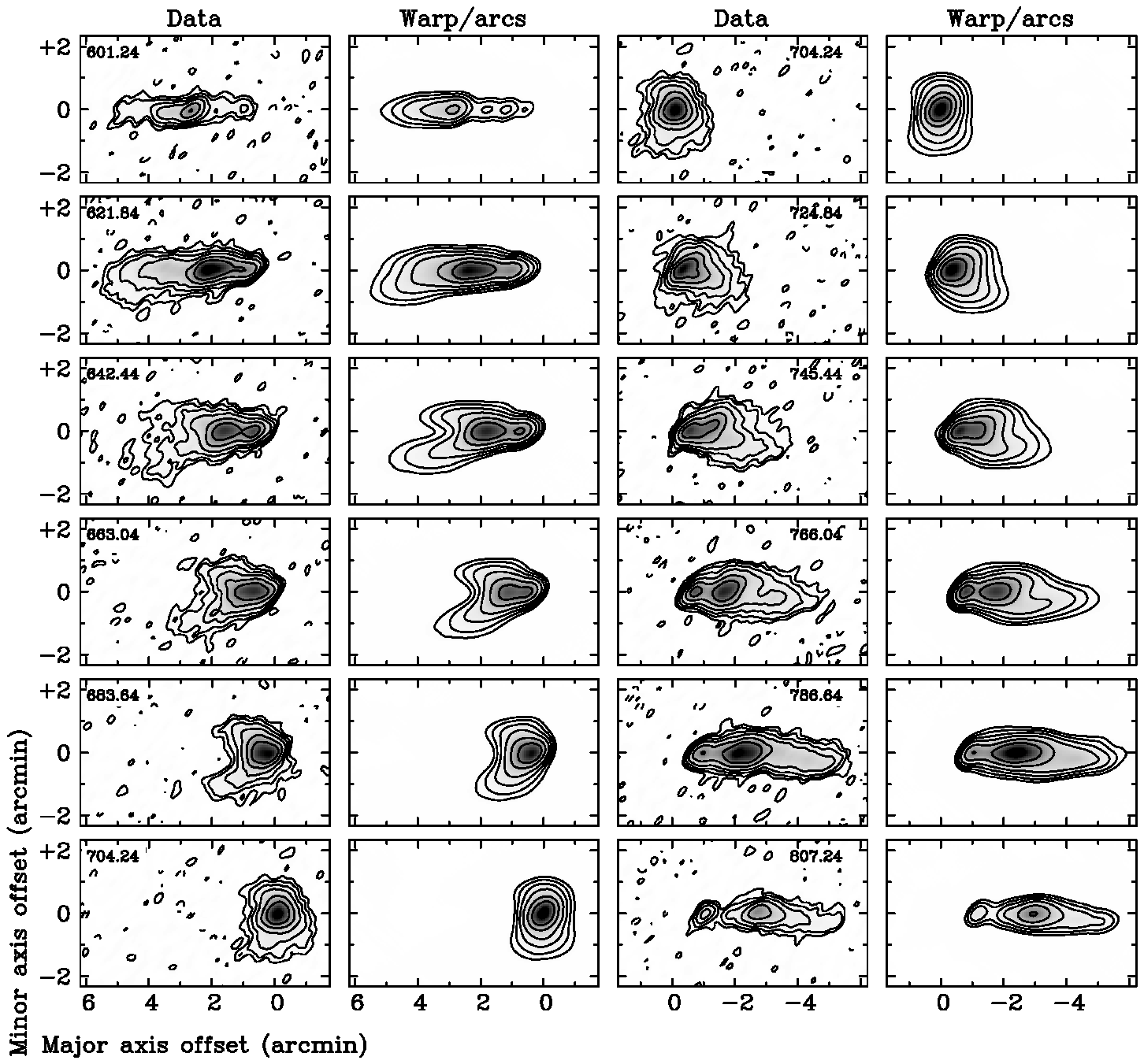}
\caption{UGC 2082: Representative channel maps in the rotated frame for the data
  and the best model (``Warp/Arcs''). Contour levels are the
  same as in Figure \ref{fig5a}.}
\label{fig4a}
\end{figure*}

\begin{figure*}
\centering
\includegraphics[width= 16 cm]{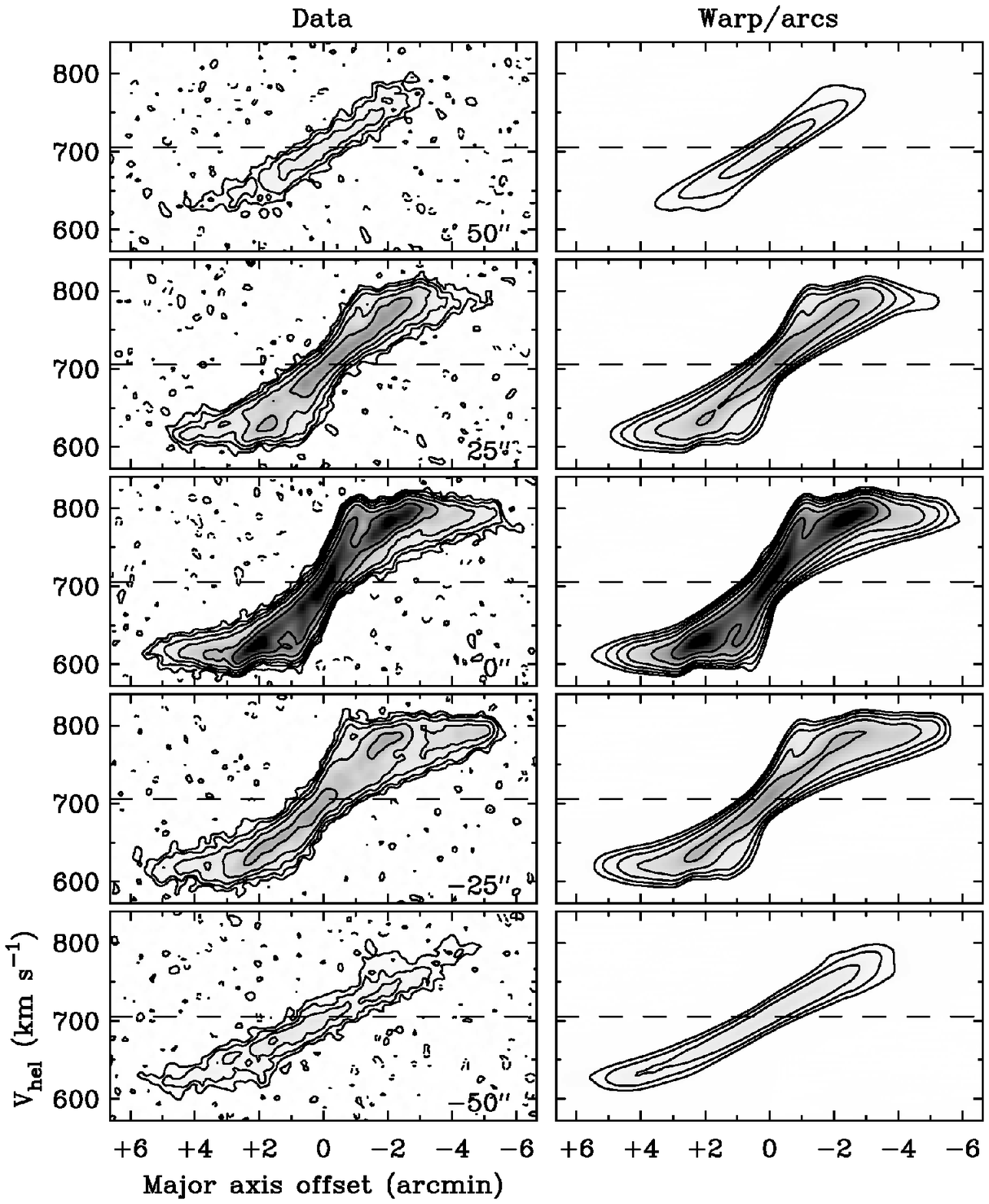}
\caption{UGC 2082: Representative PV-diagrams parallel to the major axis at different minor axis offsets (shown
in each panel; positive offsets are to the north in the rotated frame) for
  the data (left) and the best model (right). Dashed lines in the data panels indicate the systemic velocity. Contour levels are the
  same as in Figure \ref{fig5a}.}
\label{fig6}
\end{figure*}

We turn to the isolated, low SFR edge-on UGC 2082. Contours of \HI\ emission from the full-resolution cube are 
shown in Figure \ref{fig0}. From an initial examination presented in \cite{2011A&A...526A.118H}, it was already clear that the HI emission was extended in the
minor axis direction, and a position-velocity diagram parallel to the
minor axis suggested a line-of-sight warp is largely
responsible.  The zeroth moment map also suggested a warp component
across the line of sight.  The minor axis extent was noted to be
greater on the approaching (SE) side of the galaxy, particularly
evident on the S quadrant on that side.  Some high latitude extensions
were also seen on the NE side of the disk. The total \HI\  mass detected is $2.54 \times 10^9$ \msun, which agrees well
with that measured from single-dish observations of $2.4 \times 10^9$
\msun\  \citep{2005ApJS..160..149S}.  Small-scale, isolated features at the first contour level in Figure \ref{fig0} represent noise peaks and are not significant.\\
\indent Here we attempt to understand the distribution and kinematics of the
\HI\ by modeling the emission in the full resolution cube. Once more, the modeling
essentially follows the standard techniques used for HALOGAS galaxies,
using primarily \tiri\ to create
model data cubes, although some details vary from the procedure used for NGC 5023.  Initial estimates for the systemic velocity and
dynamical center, and dependence of the major axis position angle (PA)
on radius were made from the {\sc gipsy} program {\sc rotcur}.  The initial estimate for the rotation curve and radial
surface density profile is derived from an initial automated fit to
the data cube using \tiri.  Inclinations near 90$^{\circ}$ are
considered initially.  The vertical density dependence of modeled
layers is described by an exponential.  A velocity dispersion is also
specified.\\
\indent No subsequent automated fitting was
done for this galaxy.  These parameters may all be manually refined by comparison of models
and data through average profiles parallel to the minor axis, zeroth
moment maps, representative channel maps, and PV-diagrams parallel to the minor and major axis,
presented in Figures \ref{fig1}-\ref{fig6}.  Many models can be
rejected using a minimum of such figures -- zeroth moment maps and PV-diagrams parallel to the minor axis most clearly reveal their deficiencies -- hence those models
will only be shown in those figures for simplicity.  We work in a
coordinate system where the galaxy has been rotated clockwise by
42$^{\circ}$ so that the major axis is horizontal in the moment and
channel maps, with the approaching side on the left, and the NW side
referred to as the N side in what follows.  Given the clear
asymmetries in this galaxy, all four quadrants were modeled
independently unless otherwise specified.  There are only slight
deviations in the rotation curves and radial surface density profiles
among the models considered; hence we will present these only for the
preferred model.  The systemic velocity is 705 km s$^{-1}$, the
dynamical center is at R.A. $2^{\rm h} 36^{\rm m} 16.6^{\rm s}$,
Dec. $25^{\circ} 25\arcmin\  20\arcsec$, and the velocity dispersion (found to match
the falloff of the contours on the terminal side of the PV-diagram along the major axis in Figure \ref{fig6}) is $10$ km s$^{-1}$ in all models (with the exception of a few central rings in the final model, to be discussed later), with no significant change with height.\\
\indent The zeroth moment map (Figure \ref{fig2}) of the full-resolution cube suggests a warp
component across the line of sight.  The aforementioned minor axis extension is still
most prominent in the southern approaching (left) quadrant  but some low
level, extended emission is apparent in the northern approaching and
southern receding quadrants as well.  The PV-diagrams parallel to the minor axis (Figures \ref{fig5a} + \ref{fig5b}) show that this
emission is found near the systemic (top of each panel on the
approaching side, bottom on the receding side) end, at distances
within about 4\arcmin\   along the major axis from the center, with velocities
moving towards systemic with increasing minor axis offset, while being
rather narrow in velocity extent.  As mentioned above, this appearance
suggests a warp component along the line of sight is primarily
responsible, although we will see that such a model cannot explain the
high latitude emission completely.\\
\indent All models feature a warp across the line of sight.
Initially, models featuring (a) a single, unflared disk, (b) two such
disks with differing scale heights (to search for a thick disk), (c) a single flared disk, and (d) a single disk with the addition of a line-of-sight warp component were created and optimized by eye
to fit the data.\\
\subsection{Single Disk Models} \label{U20821d}
For the single disk case, separate models were created for the
approaching and receding sides rather than for all four quadrants.
Only combinations of inclination ($86-90^{\circ}$ and scale height
($11.5-15$\arcsec or $800-1050$ pc) that provided a good match to the averaged minor axis
profile (Figure \ref{fig1}) were considered.  This model can be quickly ruled out from its
appearance in the PV-diagrams parallel to the minor axis in Figures \ref{fig5a} + \ref{fig5b}.  In panels for major axis offsets at 3\arcmin\ 
or less from the center, compared to the data, there is too great a
vertical extent on the terminal end relative to the systemic end.  No
combination of scale height and inclination alleviates this problem
while still providing a good match to the minor axis profile.  The model
also cannot reproduce the shape of the high latitude extensions.\\
\subsection{Two-disk Models} \label{U20822d}
The two-disk models begin with the single-disk models and add a thicker
disk, constrained to have the same parameters as the thin disk but
with a larger scale height.  Various scale heights were experimented
with, but these models can quickly be ruled out by inspection of the
PV-diagrams parallel to the minor axis (Figures \ref{fig5a} and \ref{fig5b}).  Particularly
evident in the $2-3\arcmin\ $ panels, much of the high latitude emission occurs at velocities
close to the terminal velocity end, whereas more is needed at the
systemic end.  No variation of the parameters of this model will
remove this generic deficiency.  Given these failures, no attempt was
made to optimize these models for each quadrant independently.\\
\subsection{Flare Models} \label{U2082f} In the flare models, the scale height of the unflared inner disk is
6\arcsec\ and the inclination is $84.5^{\circ}$ in all rings.  All four
quadrants are modeled independently, but only results for the south
side will be shown, as this is sufficient to demonstrate the
inadequacies of these models.  The flare begins at $R=3\arcmin\ $ (approximately 
$R_{25}$) in all
quadrants, but the increase of the scale height with radius varies
among quadrants, with the most extreme rise in the southern
approaching quadrant, where it rises linearly to 56\arcsec\ at $R=6\arcmin\ $, and
the least extreme in the northern receding quadrant, where the scale
height increases to 12\arcsec\ at $R=3.7\arcmin\ $ and stays constant thereafter.
These dependencies allow the minor axis profile to be optimally fit, but
the generic failure of these models does not depend on the details of
the inner radius of the flare or its severity.  These failures are
demonstrated by the zeroth moment map and the PV-diagrams parallel to the minor axis.  In the
former (Figure \ref{fig2}), the models generally fail to reproduce the tapering of
emission with increasing major axis offset.  In the latter (Figures \ref{fig5a} and
\ref{fig5b}), they are a
particularly bad match to the data in the $3-5\arcmin\ $ panels, producing
high latitude emission with too broad a velocity width at velocities
too close to the terminal end in the $3\arcmin\ $ panels, and overestimating
the minor axis extent in the $4-5\arcmin\ $ panels.\\
\subsection{Line-of-sight Warp Models} \label{U2082w} The model with a warp along the line of sight is more successful.
Different scale heights and dependences of the inclination with radius
were experimented with until an optimal match was found to the channel maps and PV diagrams while still fitting the minor axis profile well.  The
inclination of the unwarped disk is $84.5-85^{\circ}$, and the
line-of-sight warp begins at about $R=200\arcsec$ (again close to $R_{25}$).
The warping is initially to smaller inclinations, reaching 
$78^{\circ} - 79^{\circ}$, depending on the quadrant, but on the
receding side improvement was found by again increasing the
inclination back to near $88^{\circ}$ at major axis offsets $>4.5\arcmin\ $
[otherwise the width of the emission in the channel maps (Figure \ref{fig4a})
at these offsets is clearly overestimated].  We do not discuss the run of
inclination in this model any further as it is not the optimal model,
although a simplified version of it is incorporated in our preferred
model which is discussed in more detail later.  Despite the
improvement over the other models, this model suffers from several
inadequacies.  First, as for all models, different parameters in all
four quadrants are necessary to fit the high latitude extensions, as
is clear in the PV-diagrams parallel to the minor axis (Figures \ref{fig5a}
and \ref{fig5b}).  This raises doubts as to whether a
line-of-sight warp component by itself is well motivated to explain
these features.  Second, the shape of the PV-diagrams parallel to the minor axis on the receding
side in the $1-3\arcmin$ panels (Figure \ref{fig5b}) is still not well reproduced.  In the $1\arcmin$
panel, the contours on the N side for the data are much straighter
than in the model, and the high latitude extension on the S side is
poorly matched.  In the $2\arcmin$ panel, the shape of this high latitude
extension is again poorly reproduced (also true in the $3\arcmin$ panel),
and emission is underestimated at velocities $>750$ km s$^{-1}$ and
minor axis offsets $>30$\arcsec\ on the north side.  Third, for the
approaching side PV-diagrams parallel to the minor axis (Figure \ref{fig5a}), especially at $2\arcmin$, $3\arcmin$ and $4\arcmin$
offsets, the southern minor axis extensions are also poorly matched,
with the data showing more extended low level emission.  These
inadequacies are also reflected in model channel maps (not shown).\\
\subsection{Modeling the Major Asymmetries} \label{U2082asym}
\begin{figure}
\centering
\includegraphics[width= 8 cm]{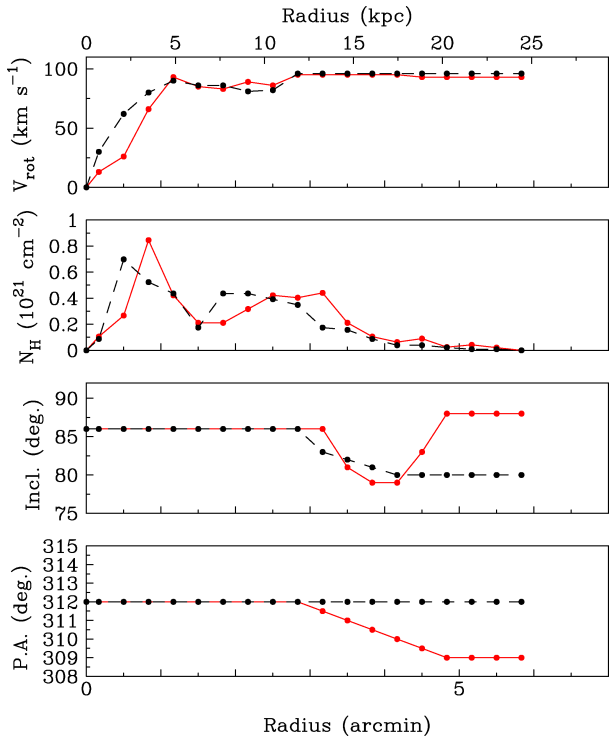}
\caption{UGC 2082: Parameters for the best base warp model for the approaching (dashed black lines) and receding (solid red lines) sides.  From top to bottom
  are shown the rotation curve, column density perpendicular to the
  disk, inclination and PA vs. radius.}
\label{fig7}
\end{figure}

\begin{figure}
\centering
\includegraphics[width= 8 cm]{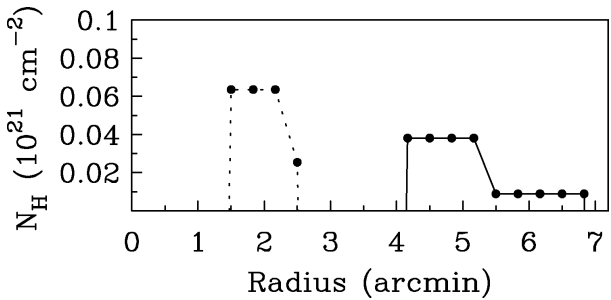}
\caption{UGC 2082: Column density perpendicular to the disk for the N (dotted) and
S (solid) arcs vs. radius.}
\label{fig8}
\end{figure}

\begin{figure*}
\centering
\includegraphics[width= 16 cm]{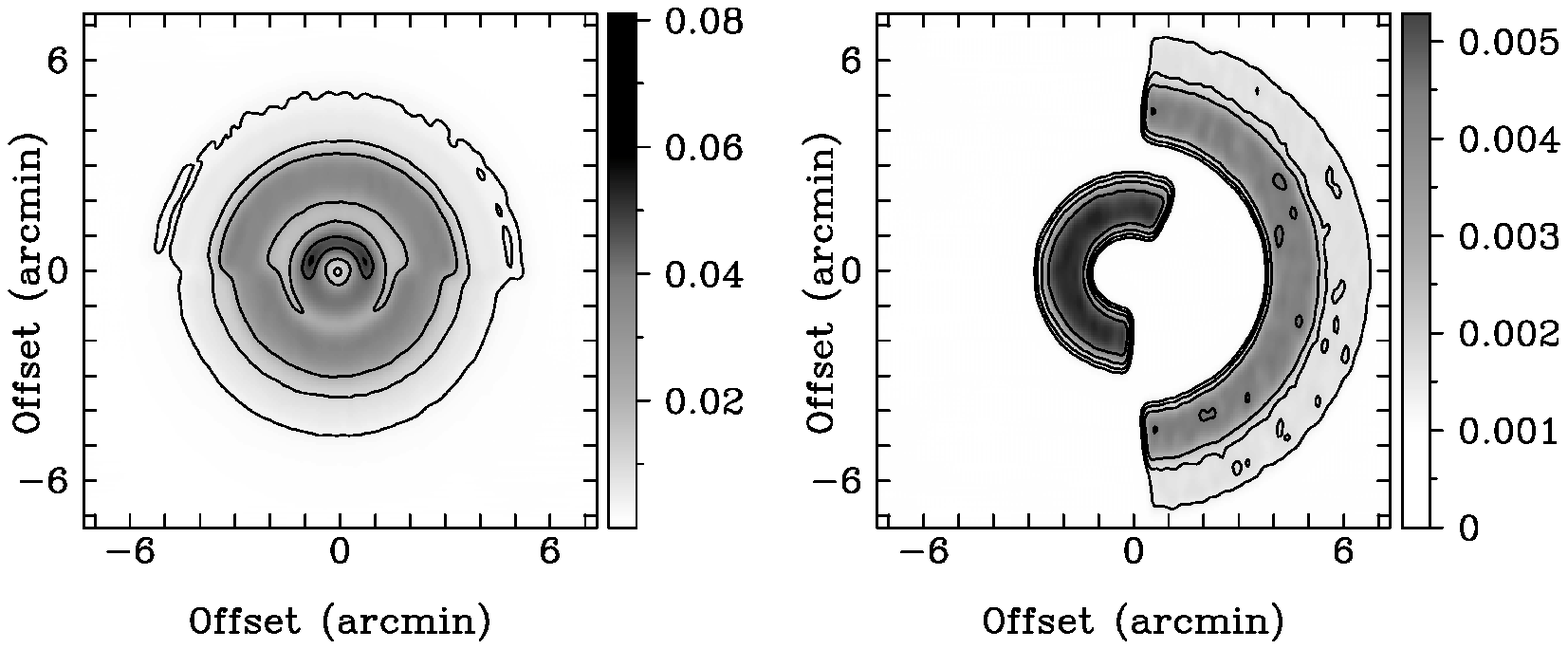}
\caption{UGC 2082: The column density of the base warp model (left) and arcs (right)
  from a face-on perspective (with respect to the unwarped part of the
  base warp model).  The models are observed from the right (the S side in
  the rotated frame of Figure \ref{fig2} is assumed to be the near side). Contour levels are $2^n \times 5.1 \times 10^{19}$\ cm$^{-2}$, $n=0-4$ for the disk, and $2^n \times 5.1 \times 10^{18}$\ cm$^{-2}$, $n=0-3$ for the arcs.}
\label{fig9}
\end{figure*}

\begin{figure*}
\centering
\includegraphics[width= 16 cm]{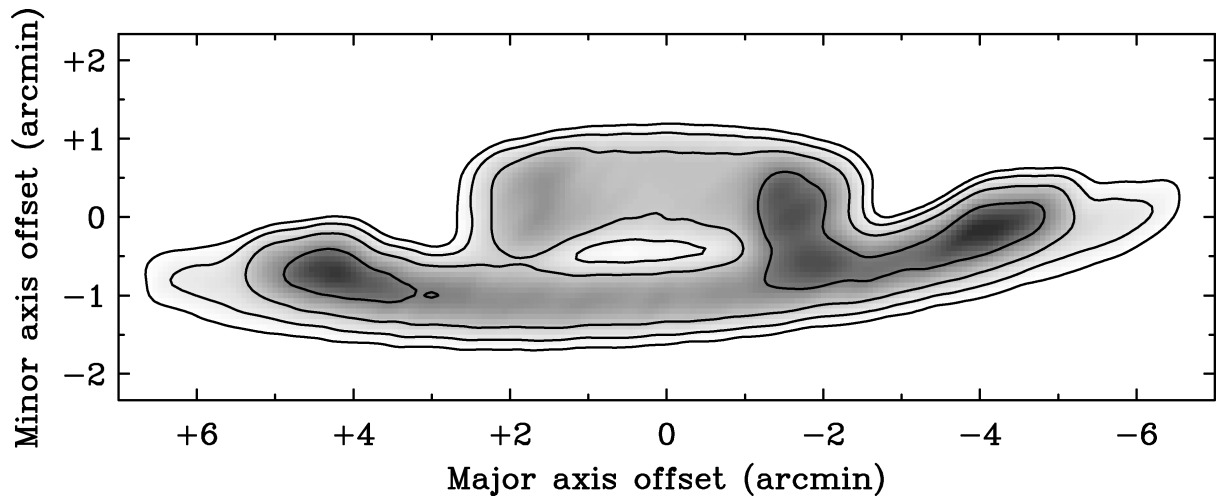}
\caption{The column density of the arcs viewed edge-on.  Contour levels are
$2^n \times  1.3 \times 10^{19}$\ cm$^{-2}$, $n=0-3$}
\label{fig9b}
\end{figure*}

Summarizing so far, all except the warp model can be clearly rejected,
but that model cannot be made to reproduce the minor-axis-extended,
asymmetric emission.  Given this asymmetry, we next attempt to augment
the warp model using \tiri's ability to model rings of restricted
azimuthal (in the plane of the galaxy) extent.  The warp model we
begin with is for the N side, separately optimized for the approaching
and receding sides.  The zeroth moment map suggests that we should add
a feature that is most noticeable in the S approaching quadrant, but
we find it must also extend into the S receding quadrant.  A PA
slightly larger and inclination slightly lower than values for the
unwarped main disk, and an azimuth range that concentrates it along
the southern side, yield such a feature that is most prominent in this
quadrant.  The arc must be placed at relatively large radii, as
the PV-diagrams parallel to the minor axis (Figures \ref{fig5a} and
\ref{fig5b}) show that this extension is concentrated at the systemic
end.  We experimented with radial density profiles, as well as
(assumed not to vary among rings in the structure) rotation
velocities, inclinations, position angles and vertical extents (we
assumed a constant density with height).  A second feature was added
to the N half in order to reproduce the straightness of the contours
in the $1\arcmin\ $ panel at minor axis offsets of $0.5-1\arcmin\ $ in
Figure \ref{fig5a} to add needed emission at the terminal end at these
offsets in the $2\arcmin\ $ panels.  Parameters were varied in the
same way as for the first component to produce the best fit.  Given
the velocities and minor axis offsets at which this feature appears,
it is necessarily at lower radii than the first arc, but with a large
vertical extent.  Both arcs were assumed to be concentric with and 
have the same velocity dispersion as the main disk.\\ \indent We note
here that we also attempted to fit these features with spiral arms as
in our model of NGC 5023, but we found these models to be a poor match
to the data.\\ \indent We attempted to keep the base warp model as
symmetric as possible, but slight changes were made.  Removing the
warp across the line of sight on the approaching side resulted in
slight improvements -- hence the downward bend of the emission towards
the end of this side is solely due to the first added component.  The
warp along the line of sight is also now slightly less severe on this
side.  On the receding side, the aforementioned behavior of the warp
across and along the line of sight is still needed to reproduce the
data.  The velocity dispersion within 50'' radius from the center has
been increased to 25 km s$^{-1}$ to better match the high velocity
width in the minor axis slice (top row in Figure \ref{fig5a}) at minor
axis offsets $<30''$.  This value does not necessarily represent an
actual increased dispersion -- it could reflect poorly resolved
non-circular motions near the center of the galaxy. No other variations in velocity dispersion were found to be necessary to model, and no discrepancies between data and models indicated that radial motions would improve the fit.  \\ 
\indent The radial variation of parameters for the base warp model is shown in
Figure \ref{fig7}, while Figure \ref{fig8} shows the radial column density
profiles for the two arcs (the deviations from a single column density for each arc do provide improvement to the quality of the fit).
Table \ref{tab1} lists the other parameters for the arcs.  The channel maps
and PV-diagrams parallel to the major axis for the data
and this model are compared in Figures \ref{fig4a} and \ref{fig6}, 
respectively.  A face-on
view of the surface density distribution of the base warp model and
the two arcs is shown in Figure \ref{fig9} (note the unavoidable
discontinuities in the base model where the approaching and receding
sides meet), and a view of the arcs alone from the observer's perspective in Figure
\ref{fig9b}.  Also note that Figure \ref{fig9} assumes that the S side
of the galaxy is the near side.  There is no prominent dust lane in
UGC 2082 which could help break this ambiguity.  The total \HI\ masses
are $6.8\times 10^7$ \msun\ and $1.3\times 10^8$ \msun\ for the N and
S arcs, respectively.\\
\indent The zeroth moment map (Figure \ref{fig2}) shows how the addition of these
arcs produces a good match to the \HI\ morphology.  The PV-diagrams parallel to the minor axis (Figures \ref{fig5a}
and \ref{fig5b}) show
how the first arc approximately reproduces the high latitude tails at
the systemic end in the $1-4\arcmin\ $ panels on both sides.  The match is not
perfect: for instance, on the approaching side, the emission in the
data is slightly more extended along the minor axis, and the slope of
the extension in the $4\arcmin\ $ panel on this side and the $2\arcmin\ $ panel on the
receding side is shallower than in the model.  The addition of the
second arc has largely solved the aforementioned problems in this
figure that motivated its inclusion.  The channel maps (Figure \ref{fig4a}) 
and PV-diagrams parallel to the major axis (Figure \ref{fig6})
also show excellent agreement, although a few isolated discrepancies
are clearly noticeable.\\
\indent When fitting by eye, it is difficult to assess the uniqueness of the arc
parameters, but, as we did for NGC 5023, we can make estimates by varying parameters in turn while
keeping the others fixed, and judging when the models cease to be good fits
to the data.  The error bars in Table \ref{tab1} result from this exercise.
The velocity dispersion of the S arc is uncertain by about 2 km s$^{-1}$, but
that of the N arc is poorly constrained.\\
\indent Following the discussion for NGC 5023, by including the arcs the number of free parameters in the model is not increased greatly.  Apart from the
six parameters listed in Table \ref{tab1} for each arc, no more than two
distinct column densities per ring were experimented with (as reflected in
\ref{fig8}, although variations in the central radius and width of the rings were also examined.  Hence, effectively only 20 free parameters were added to the base warp model, which itself has almost 100 free parameters.\\
\indent Beyond the features listed in Table {\ref{tab2}, no additional small-scale or low-level emission was found in an
examination of the full-resolution cube.  The favored model provides a
reasonable fit to the 30\arcsec\ cube and no significant modification of the
parameters is required.\\
\indent Finally, we note that no lag is required in the fitting.  There
is no residual characteristic lag signature in Figures \ref{fig5a} and 
\ref{fig5b} of the type seen in NGC 5023 that would necessitate its inclusion.
It should be recognized, however, that the linear resolution perpendicular to the disk ($\sim$ 1.5 kpc) is worse than in other edge-ons ($350-900$ pc; \citet{Oosterloo2007,Kamphuis2011,Zschaechner2011,Zschaechner2012}) where lags have been measured.  It is therefore possible that higher resolution mapping could reveal a lag relatively close to the midplane.\\
 \begin{table}
    \centering
    \begin{tabular}{@{} llll @{}} 
       \hline
       \hline
& Disk & N arc  & S arc \\
       \hline
Center ($\alpha$ J2000) & 2$^{{\rm h}}$  36 $^{{\rm m}}$16.6$^{{\rm s}}$ & & \\
	 \hspace*{0.93cm}($\delta$ J2000) & 25$^{\circ}$ $25\arcmin\ 20\arcsec$ & &\\	
$V_{\rm sys}$ (km s$^{-1}$)&705& &\\

Incl.(deg.) & $86$ & $78 \pm 5$ & $78 \pm 1.5$ \\
PA (deg.) & $312$ & $312 \pm 3$ & $316 \pm 2$\\
$V_{rot}$ (km s$_{-1}$) & 95 & $85\pm 5$ & $85 \pm 5$ \\
$\Delta z$ (arcsec) & & $40\pm 5$ & $20 \pm 5$\\
$AZ_0$ (deg.) & & $255\pm 10$ & $90 \pm 10$\\
$\Delta AZ$ (deg.) & & $205\pm 10$ & $170 \pm 10$\\
Mass (\msun) & & $6.8 \pm 1.7 \times 10^7$ & $1.3 \pm 0.3 \times 10^8$ \\
Mass (\%) & & $2.7 \pm 0.7$ & $5.1 \pm 1.2$ \\
\hline
    \end{tabular}
    \caption{UGC 2082: Parameters for the Disk and Two Arcs. The disk inclination and position angle are for the unwarped region.  The disk and arcs have the same dynamical center and systemic velocity.  The disk rotation speed is an average value for the flat part of the rotation curve beginning at 3 \arcmin radius. $\Delta z$ is the vertical extent of each arc.  $AZ_0$ is
the central azimuth in the plane of the galaxy, with 0$^{\circ}$ being the
approaching side of the major axis.  $\Delta AZ$ is the azimuthal extent of
each arc in the plane of the galaxy.  The final row gives the percent of the \HI\ mass of the galaxy accounted for by each arc.}
    \label{tab1}
 \end{table}
 \section{Discussion}\label{disc}
 \subsection{NGC 5023 and spiral features}
Through extensive modeling of the \HI\ observations of NGC 5023 we have shown that this galaxy most likely contains spiral arms, but definitely some kind of non-axisymmetric gas distribution. This is the first time that such models have been created for  an edge-on galaxy in order to investigate its effect on the observed vertical gradients in rotational velocity. We have seen that in a ``classical'' analysis of the lag in NGC 5023 the vertical gradient would be  $\sim$ -15 $\pm$ 3  km s$^{-1}$ kpc $^{-1}$ up to a radius of 133\arcsec (4.3 kpc) and  then decline to 0 at 304\arcsec  (9.7 kpc). As this best fit model is without global line of sight effect, i.e. a line of sight warp or flare,  this vertical gradient can be considered an absolute upper limit on the vertical gradient in NGC 5023. However, if we include non-axisymmetric features in the form of spiral arms the vertical gradient derived from the best fit model would be lower  (-9.4 $\pm$ 3.8  km s$^{-1}$ kpc $^{-1}$) and constant with radius. Although there are indications in the data that a radial variation is present, including such a variation no longer improves the model fit significantly.\\ 
 \indent The fact that non-cylindrical distributions of the gas above the disk  could severely affect the measurement of the vertical gradient has been known for a long time. However, until now it was impossible to accurately model such a distribution. With the improved capabilities {\sc TiRiFiC} we can now construct models that match the data to an unprecedented level of accuracy. Even though this increases the range of possible models that can be investigated it also introduces a large set of new free parameters to the models. This means that any additions to the ``classical'' two-disk cylindrical symmetric models should be motivated physically as well as improve the fit to the data significantly before they can be accepted as the better model. \\
 \indent Although the spiral arms in our model of NGC 5023 may have a slightly different scale height than the disk, their thickness is not enough to explain the individual features described in section \ref{N5023ind} (see Figure \ref{mom0Model}). Unfortunately the features are closely connected with the disk making it impossible to acquire reliable masses. Therefore a simple comparison between the energy required to expel them from the disk and the SFR becomes impossible. However, Figure \ref{Face-Onview} shows that the arms extend the longest in a single line of sight (the x-axis in Figure \ref{Face-Onview}) around 2\arcmin offset along the major axis (y-axis), very close to the locations of the features (Table \ref{tab2}). This and their proximity to the disk seem to point at an internal origin.\\
 \indent  The difference between a lag obtained from a ``classical'' model and the lag in our Spiral Arm model is driven by the presence of the arms in the line of sight warp. It is therefore questionable whether  such asymmetries will have an effect in galaxies that, unlike NGC 5023, contain a significant thick disk or are perfectly edge-on with no pronounced asymmetries above the mid-plane. If the gas is pumped out of the plane by supernovae, it is logical to assume that the extra-planar gas close to the disk is densest above an area near the arms thus causing similar asymmetries above the plane. However, is there any observational evidence supporting such a picture? In the non-edge on cases where anomalous gas has been detected the spiral arms cannot be identified in the anomalous component, either due to resolution problems or because a grand design spiral is absent. However, in the case of NGC 6946 the anomalous gas is clearly located towards the star forming regions of the galaxy \citep{2008A&A...490..555B} and at the highest velocity offsets ($|\Delta {\rm v}|\geq 80$ km s$^{-1}$, their Figure 9) a clear asymmetry in the distribution is visible. An asymmetry can also be tentatively seen in the anomalous gas distribution of NGC 4559 \citep[Figure 6]{2005A&A...439..947B}. However, the anomalous gas in NGC 2403 appears to form a smooth circular disk \citep{2002AJ....123.3124F}. In the case of NGC 891 the extra-planar \Halpha\ distribution is likely to be correlated with the spiral arms \citep{Kamphuis2007b}. Even though, the thick \HI\ disk of this galaxy is so massive (M$_{\rm H\textsc{i}\ thick\ disk}$= 0.3 M$_{\rm H\textsc{i}\ total}$, \cite{Oosterloo2007})   that   it is unlikely that all the neutral gas is located above the arms, a non-cylindrically symmetric over-density could suffice to significantly lower the derived lag as the line of nodes becomes more and more devoid of gas. \\
 \indent Although spiral arms are well motivated, there are still many caveats with the best fit model of NGC 5023 model. The only reason for  assuming a two-arm grand design spiral is that it is the simplest spiral arm model, but that does not mean that any other model with non-cylindrically symmetric features,  e.g. a four-arm model or a flocculent spiral, cannot fit the data. Also, in our Spiral Arm model for NGC 5023 {\sc TiRiFiC} has organized the arms in such a way that the effect on the observed vertical gradient is maximized, i.e. the over density is as far from the line of nodes as possible (See Figure \ref{Face-Onview}), thus minimizing the gradient. On the other hand, there is no real need to organize the gas in spiral arms. All that is required in order to simulate the lag is that the distance between the line of nodes and the gas increases as a function of distance to the central plane -- a likely situation, as only in the case  where the gas above the plane is, up to its maximum vertical extent, on the line of nodes would there be no effect on the derived rotational velocities.\\
 \indent Our models also do not explore radial motions or variations in dispersion. such variations are observed in other galaxies \citep{2002AJ....123.3124F, 2005A&A...439..947B,Oosterloo2007}. A variation in the dispersion seems unlikely as a comparison between the final model and the data does not show any major discrepancy which could relate to a variation in dispersion with height (See Figures \ref{YVLocal} and \ref{5023XV}) However, radial motions are often associated with spiral arms and could help to reproduce the boxiness of the central channel maps (Figure \ref{5023Chans}, bottom left/top right panel). There are several reason why radial motions are not included in the models, first of all they would present another set of free parameters. Due to the edge-on orientation of the galaxy the location of the motions would be ill constrained. As they would only help to fix some minimal discrepancies between the data and the final model in the central channels the improvement of such a model would be minimal at best. We could constrain the radial motions by merely allowing their presence in the spiral arms of the final model, however this model already has a high number of free parameters. Additionally, as discussed at the start of the previous paragraph, it is merely an example of a realistic gas distribution and therefore it would be very speculative to constrain radial motions to the arms in the model.\\  
\indent If it turns out that many of the observed vertical gradients are overestimated in current tilted ring models, this could have a profound impact on the theory of disk-halo interactions. The velocity difference between ballistic models and observed lags \citep{2008MNRAS.386..935F} could become less pronounced. However, in the case of NGC 5023 the vertical gradient is affected due to an overdensity in the line of sight warp. Do asymmetries in diffuse thick disks have the same pronounced effect on the vertical gradient? In how many nearly edge-on galaxies is the line-of-sight warp discarded due to the omission of non-cylindrically symmetric modelling? To answer these question the addition of spiral arms to the tilted ring models should be investigated for previous and forthcoming results.\\  
\subsection{UGC 2082 and arc features}
We turn our attention to the origin of the two modeled arcs in UGC
2082.  We first reiterate that, unlike for NGC 5023, models that
attempted to fit these features with spiral arms were a poor match to
the data, weakening the possible interpretation of the emission as
disk-halo cycled gas above spiral arms in this case.  We note that UGC
2082 has a similarly low global star formation rate to NGC 5023, but a
disk of much larger radial extent, and therefore we may expect weaker disk-halo
cycling.

The southern arc is at large radii and for this reason too is unlikely
to represent disk-halo cycled gas.  UGC 2082 is also a very isolated
galaxy: it is not a member of any group optically identified by
\citet{2000ApJ...543..178G}, while the closest galaxy with detected
\HI\ in a recent search with the Arecibo telescope is at a projected
distance of 279 kpc \citep{2011AAS...21724606T}, scaled to our
distance.  Therefore, the arc is unlikely to be due to a tidal
interaction.  Even if the arc represented an asymmetric warp, these
tend to occur in rich environments \citep*{2002A&A...394..769G}, while
the same authors find that lopsidedness is associated with nearby
companions.  Therefore, it seems that the arc is a candidate for an
accretion event, although this conjecture is based only on the lack of
a plausible alternative explanation.

Since the northern arc is above the inner disk, one may yet ask
whether it could be a result of a disk-halo flow, despite the low star
formation activity, and keeping in mind that the extent of the
structure is much larger than one would expect from a single kpc-scale
supershell creation event.  Nevertheless, one can estimate the number
of supernovae required to lift such a mass of gas to a certain height
above the disk under several assumptions.

First, the potential energy of a mass of gas at a height z above a
disk, in the case of an isothermal sheet, is given by \citet{1997AJ....114.2463H}:

$
\Omega = 10^{52} \hspace{0.15 cm} {\rm erg} \left({M_{\rm cloud}\over 10^5 {\rm \hspace{0.15 cm}  M_{\sun}}}\right)\\
\hspace*{1.0 cm}\times\left({z_0\over 700 {\rm \hspace{0.15 cm} pc}}\right)^2\left({\rho_0\over 0.185 \hspace{0.15 cm} {\rm  M}_{\sun}\hspace{0.15 cm}  {\rm  pc}^{-3}}\right){\rm ln[cosh(z/z_0)]}
$

where $M$ is the mass of the gas and $z$ is its height above the
plane, $z_0$ is the mass scale height of the disk, and $\rho_0$ is the
mass density in the midplane below the feature.  We assume that the
disk potential is the only relevant contributor to the potential here.
For the disk scale height, we can use the relation between stellar
scale height and rotation speed found by \citet{1999A&A...352..129V}
to infer a value of about 300 pc.  We assume this is a reasonable
approximation for $z_0$. For the mass density, we begin with the value adopted by \citet{1997AJ....114.2463H} for NGC 891.  We then assume that the total stellar masses follow the stellar-mass Tully Fisher relation of \citet*{2005MNRAS.358..503K}, to find that the stellar mass of UGC 2082 should be 0.035 that of NGC 891.  We next assume that a rough, average stellar density ratio of the two galaxies can be determined from this ratio by scaling by the product of the K-band disk area and scale height.  The disk areas are taken from the 2MASS survey \citep{2006AJ....131.1163S}, while the scale heights are assumed to follow the correlation with rotation speed from \citet{1999A&A...352..129V}.  We thus estimate a midplane density of 0.04 \msun/pc$^3$ for UGC 2082. This is of course only a rough average for the whole disk.  Using half
of the vertical extent of the arc for $z$ (1.4 kpc), we find a
potential energy of about $1.0 \times 10^{54}$ erg.  Assuming
$10^{51}$ ergs per type II supernova, and if about 10\% of this energy
goes into the kinetic energy of the gas \citep{1988ApJ...324..776M},
then $10^4$ SN are required.  If the vertical velocity of the gas is
50 km s$^{-1}$ [conservatively at the slow end of expansion velocities
  in the superbubble model of \citet*{1989ApJ...337..141M}], then the
timescale to lift the gas to this height is $3 \times 10^7$ years, and
the SN rate then required is $3 \times 10^{-4} {\rm yr}^{-1}$.  Using
the relation between the Type II SN rate and the SFR from
\citet{2011MNRAS.412.1508M}, the total Type II SN rate in UGC 2082 is
$4\times 10^{-4} {\rm yr}^{-1}$.  Thus, it seems, essentially all of
the SNe at this rate would need to occur in this small part of the
disk in order to raise this gas from the plane.  Although the
uncertainties are many, it does therefore appear unlikely that the arc
can be explained as a result of disk-halo cycling.  Therefore, an
accetion origin cannot be ruled out.
  
 We searched our HALOSTARS image for optical emission that might be associated with the arcs.  A faint galaxy of detectable angular extent roughly 20\arcsec $\times $10\arcsec at RA 02$^{\rm h}$ 36$^{\rm m}$  24.6$^{\rm s}$ , Dec 25$^{\circ}$ 22\arcmin 40.4\arcsec is coincident with the southern arc.  However, this may be a background galaxy. The Sloan Digital Sky Survey provides no additional information as it did not cover this area, while there is no coincident radio source in the NRAO VLA Sky Survey \citep{1998AJ....115.1693C}, or in GALEX NUV and FUV (1266 and 1574 s exposure times, respectively) images that we have obtained as
part of the HALOGAS project.
\section{Summary \& Conclusions}\label{sum}
We have investigated deep \HI\ observations of the small galaxies UGC 2082 and NGC 5023. These observations were obtained as a part of the HALOGAS Survey and wil contribute to our understanding of disk-halo interactions and cold gas accretion in the local universe.  We have constructed detailed tilted ring models of these galaxies and compared these to the full 3D information available in the observed \HI\ cubes. Through a new functionality in \tiri\ that allows the addition of localized overdensities to the models we have, for the first time, investigated the effect of non-cylindrically symmetric distributions of the neutral hydrogen on the observed data.\\
\indent In the case of NGC 5023 a ``classical'' analysis, without  non-cylindrically symmetric components, would lead us to conclude that this galaxy has no line-of-sight warp nor a flare but merely a single disk with a scale height of 10.5\arcsec (0.34 kpc) with a radially varying lag. This lag would be $\sim$ -14.9 km s$^{-1}$ kpc$^{-1}$ in the center of the galaxy  and drop to 0 km s$^{-1}$ kpc$^{-1}$ between 133\arcsec (4.3 kpc) and 304\arcsec  (9.7 kpc) radial offset. As usual for this kind of modeling the two sides of the disk were treated independently. However, if we include non-cylindrically symmetric distributions the fit to the data is significantly improved provided that the disk is warped in the line of sight. Compared to the ``classical'' analysis the scale height of the main disk is somewhat lower  (9\arcsec, 0.29 kpc),  the derived lag is significantly lowered and radial variation of the lag no longer improves the fit to the data. Additionally the underlying main disk is now almost fully cylindrically symmetric and all the asymmetry is in the warp and the localized overdensities.

No lag is required for UGC 2082, but two arcs (features of limited
extent in azimuthal angle in the disk) with a large vertical extent
and masses of order $10^8$ \msun\ each are necessary to reproduce
prominent asymmetries in the gas distribution and kinematics.  The main
disk to which they are added still requires a warp.  One arc
is at relatively large radii and features a slightly different
inclination and position angle to the main disk, while the other is
found above the inner disk.  Given the lack of nearby companions, and the insufficient star formation rate that could be responsible for these features, they may be examples of ongoing accretion.

 \section*{Acknowledgments}
P.K. would like to thank the German Humboldt Foundation for the financial support during his stay in Germany. R.J.R, R.A.M.W and M.T.P. acknowledge support from the National Science Foundation under grant AST-0908126 to R.J.R. and R.A.M.W. GG is a postdoctoral researcher of the FWO- Vlaanderen (Belgium). P.S. is a NWO/Veni fellow. The work of W.J.G.dB. was supported by the European Commission (grant FP7-PEOPLE-2012-CIG \#333939). This publication makes use of data products from the Two Micron All Sky Survey, which is a joint project of the University of Massachusetts and the Infrared Processing and Analysis Center/California Institute of Technology, funded by the National Aeronautics and Space Administration and the National Science Foundation. The Westerbork Synthesis Radio Telescope is operated by ASTRON (the Netherlands Institute for Radio Astronomy) with support from the Netherlands Foundation for Scientific Research (NWO). The authors also want to extend their thanks to the other members of the HALOGAS collaboration who have contributed with useful discussions on the paper. Finally we would like to thank the referee for carefully reading the manuscript and providing useful comments.
      
\bibliographystyle{mn2e}   
\bibliography{Kamphuis_2013_References}

\end{document}